\documentclass[]{jfm}

\usepackage{graphicx}
\usepackage{newtxtext}
\usepackage{newtxmath}
\usepackage{natbib}
\usepackage{hyperref}
\hypersetup{
    colorlinks = true,
    urlcolor   = blue,
    citecolor  = black,
}

\newcommand{\RomanNumeralCaps}[1]
\linenumbers

\title{Transition to turbulence in the wide-gap spherical Couette system}

\author{A. Barik\aff{1}
  \corresp{\email{abarik@jhu.edu}},
  S. A. Triana\aff{2},
  M. Hoff\aff{3},
  \and J. Wicht\aff{4}}

\affiliation{\aff{1}Johns Hopkins University, Baltimore, USA
\aff{2}Royal Observatory of Belgium, Brussels, Belgium
\aff{3}Brandenburgische Technische Universit\"{a}t Cottbus - Senftenberg, Cottbus, Germany
\aff{4}Max-Planck-Institut f\"{u}r Sonnensystemforschung, G\"{o}ttingen, Germany
}

\newcommand{\Ro}{\Delta\Omega/\Omega}

\newcommand{\tw}{\textwidth}
\newcommand{\U}{\boldsymbol{u}}
\newcommand{\Q}{\boldsymbol{Q}}
\newcommand{\zh}{\hat{\boldsymbol{z}}}

\newcommand{\ci}{\mathrm{i}}

\begin{document}
\maketitle

\begin{abstract}
The spherical Couette system consists of two differentially rotating concentric spheres with a fluid filled in between. We study a regime where the outer sphere is rotating rapidly enough so that the Coriolis force is important and the inner sphere is rotating either slower or in the opposite direction with respect to the outer sphere. We numerically study the sudden transition to turbulence at a critical differential rotation seen in experiments at BTU Cottbus - Senftenberg, Germany and investigate its cause. We find that the source of turbulence is the boundary layer on the inner sphere, which becomes centrifugally unstable. We show that this instability leads to generation of small scale structures which lead to turbulence in the bulk, dominated by inertial waves, a change in the force balance near the inner boundary, the formation of a mean flow through Reynolds stresses, and consequently, to an efficient angular momentum transport. We compare our findings with axisymmetric simulations and show that there are significant similarities in the nature of the flow in the turbulent regimes of full 3D and axisymmetric simulations but differences in the evolution of the instability that leads to this transition. We find that a heuristic argument based on a Reynolds number defined using the thickness of the boundary layer as a length scale helps explain the scaling law of the variation of critical differential rotation for transition to turbulence with rotation rate observed in the experiments.
\end{abstract}

\begin{keywords}

\end{keywords}

\section{Introduction}

The spherical Couette system consists of two concentric spheres differentially rotating about a common axis, with the space in between filled with a viscous fluid. The differential rotation is considered `positive' when the inner sphere rotates faster than the outer sphere and `negative' when it rotates slower or in the opposite direction as the outer sphere. Being the spherical analogue of the more well-known Taylor-Couette system \citep{Chossat94}, it is an interesting fluid dynamical system in its own right with very different instabilities.  Applications to the interiors of astrophysical bodies (e.g: planetary interiors, stellar radiative zones)  seem more obvious than in the Taylor-Couette geometry. The study of the spherical Couette system goes back to the analytical asymptotic formulation of \cite{Proudman56} for an infinitely fast rotating outer sphere and an infinitesimal differential rotation. He showed that most of the fluid differential rotation remains confined within the cylinder tangent to the inner sphere equator, known as the tangent cylinder (TC), while the fluid outside the TC co-rotates with the outer boundary. A complex nested shear layer at the TC, known as the Stewartson layer \citep{Stewartson66}, accommodates the jump in the fluid rotation rate and its derivatives. For a spherical Couette system with a wide-gap, this shear layer is the source of the first flow instabilities for a rapidly rotating outer boundary. Note that when the gap becomes narrow, the flow instabilities resemble Taylor rolls similar to the Taylor-Couette system \citep{EgbersRath95}. Instabilities of a Stewartson layer driven by differential rotation were first studied  experimentally by \cite{Hide67} for a cylindrical system with a differentially rotating disk and theoretically by \cite{Busse68}. For the case of the spherical Couette system, Stewartson layer instabilities as well as other instabilities have been extensively studied using experiments \citep[e.g.][]{Sorokin66, Munson3, EgbersRath95, Kelley2007, Kelley2010, TrianaPhD, DanZ2014, Hoff2016, Yoshikawa2023} and numerical computations \citep[e.g.][]{Munson1, Munson2,Hollerbach2003, Hollerbach2004, Hollerbach2006, Wei2008, Matsui2011, Rieutord2012, WichtJFM}. These studies have revealed a complex zoo of instabilities and have left many open questions.

Our previous study \citep[][hereafter B18]{Barik2018} and the present study are based on the experiments of \cite{Hoff2016} (hereafter H16) in a wide-gap spherical Couette set-up. Once the radius ratio of the two spheres is fixed, the system is characterized by two parameters, the Ekman number $E=\nu/\Omega_o L^2$ and the differential rotation, $\Delta\Omega/\Omega = (\Omega_i - \Omega_o)/\Omega_o$. Here, $\nu$ is the viscosity of the fluid, $L$ is the thickness of the spherical shell, and $\Omega_i$ and $\Omega_o$ denote the rotation rates of the inner and outer sphere, respectively. H16 and B18 both focused on the case when the differential rotation was negative, i.e., when the inner sphere rotated slower than or in the opposite direction compared to the outer sphere. At intermediate or low Ekman numbers ($3\times 10^{-6}\leq E \leq 10^{-4}$), as the differential rotation is made progressively more negative, the flow transitions through either four or five different hydrodynamic regimes: 
\begin{enumerate}
    \item an axisymmetric flow described by \cite{Proudman56}.
    \item The axisymmetric flow gives rise to a linear non-axisymmetric instability of the Stewartson shear layer with a fixed azimuthal wavenumber $m$.
    \item The first instability gives way to a regime with a mode with $m=1$. For a certain moderate to low range of Ekman numbers ($3\times 10^{-5}\leq E\leq 10^{-4}$), these two regimes may coincide and the first non-axisymmetric instability may occur in the form of $m=1$. 
    \item The above regime gives way to equatorially antisymmetric (EA) wave-like `inertial modes' which have been observed in several past studies \citep{Kelley2007, Kelley2010, TrianaPhD, Matsui2011,Rieutord2012,WichtJFM} and formed the focus of B18. 
    \item Finally, a sharp and sudden transition to bulk turbulence takes place at a critical negative differential rotation.
\end{enumerate}
These regimes have been observed in simulations of \cite{WichtJFM} and B18 and experiments of H16. More specifically, H16 observed that the transition to turbulence was characterized by a broadband temporal power spectra. Well-defined inertial mode peaks observed on top of this broadband spectra displayed an abrupt change in frequency right at the onset of turbulence. In addition, there was an increase in the spatial extent of the axisymmetric zonal flow and a decrease in the energy content of the inertial modes. They further observed a dependence of the critical differential rotation required for transition $|\Ro|_c$ on the Ekman number as $|\Ro|_c\sim E^{1/5}$.
In the present study we concentrate on this transition to turbulence, addressing the following questions: how does the flow behave during and beyond the transition and what causes its onset.


There have been a few other studies on turbulence in spherical Couette flow, but not for the radius ratios and parameter ranges used in this study. \cite{Belyaev1979} experimentally analysed a wide-gap spherical Couette system for a stationary outer sphere and postulated that the transition to turbulence seems to follow the scenario of \cite{Ruelle1971} but with several differences like the existence of discrete peaks on top of a continuous background power spectrum. \cite{yavorskaya1986} experimentally investigated a thin-gap system ($r_i/r_o = 0.9$) with both spheres rotating over a wide parameter range. They noticed that the transition to turbulence involves an onset of ``spatial intermittency'' in the form of small-scale structures on top of large-scale flow. \cite{Wulf99} studied two different gap widths ($r_i/r_o = 0.75$ and $0.67$) and found that the transition to turbulence was characterised by broadband temporal power spectra with some well-defined peaks.

The rest of the paper is arranged as follows. Section \ref{sec:numMethods} provides details of the formulation of the problem, a brief description of the numerical methods used for simulation and the methods used to construct spectrograms and distinguish regimes (i) through (v) mentioned above. Our results begin in section \ref{sec:spectra} with a discussion of the temporal and spatial spectra of flow and their variations. Section \ref{sec:flowAnalys} discusses our results in the physical space with an analysis of the mean zonal flow, angular momentum transport and the effect of turbulence on inertial modes. Section \ref{sec:ForceBalance} provides insight into the transition to turbulence by investigating force balances in the system. Section \ref{sec:BL} investigates the instability of the boundary layer at the inner boundary and its effects and provides a heuristic explanation of the $E^{1/5}$ scaling law obtained by H16. Finally, section \ref{sec:conclusions} discusses our main conclusions and open questions.


\section{Numerical methods}\label{sec:numMethods}

\subsection{Simulation setup}

\begin{figure}
\centering
\includegraphics[width=0.35\tw]{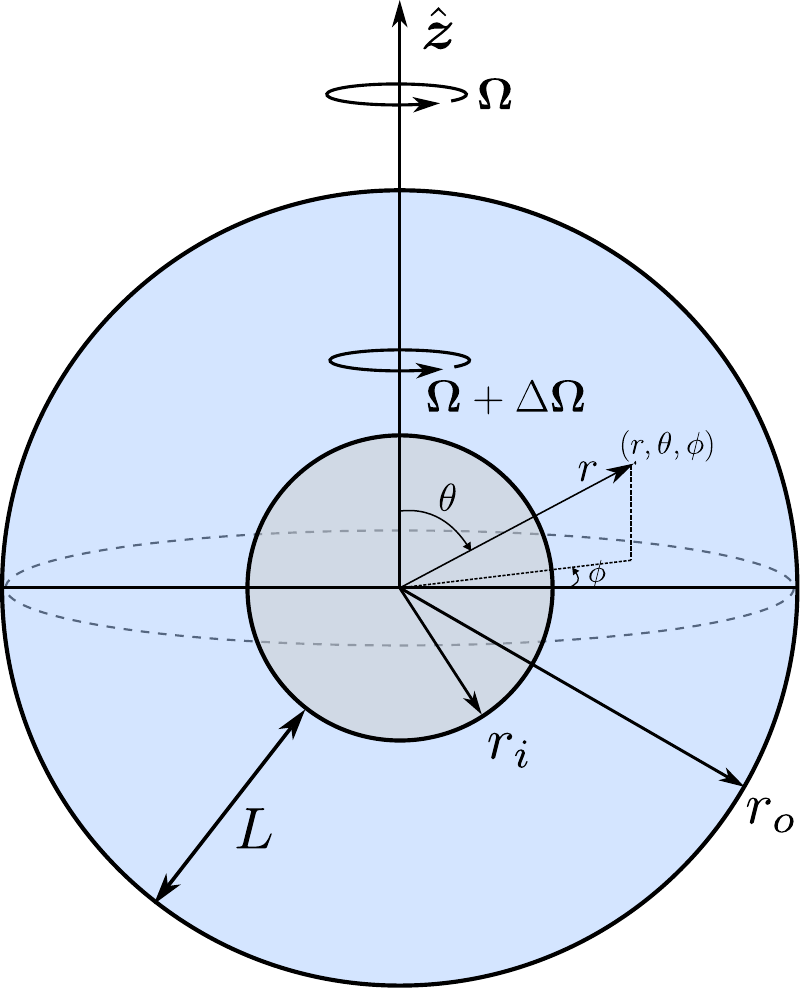}
\caption{A schematic of the spherical Couette system, indicating the rotation rates, the radii of the outer and inner boundaries and the spherical coordinate system $(r,\theta,\phi)$.}
\label{fig:sys}
\end{figure}

Let us denote the radii and dimensional rotation rates of the two coaxial spheres as $r_i$ and $\Omega_i$ for the inner sphere, and $r_o$ and $\Omega_o$ for the outer sphere, respectively. To simulate this system, we solve the Navier-Stokes and continuity equations in a reference frame rotating with the outer boundary. We use  spherical coordinates $(r,\theta,\phi)$ denoting radial distance, colatitude and longitude, respectively. We also use $s=r\sin\theta$ to denote the cylindrical radius, perpendicular to the rotation axis. The equations are non-dimensionalised using $L=r_o-r_i$ as the length scale and the viscous diffusion time $\tau = L^2/\nu$ as the time scale, where $\nu$ is the kinematic viscosity of the fluid. This gives us,

\begin{equation}\label{eqn:NS}
\displaystyle\frac{\p \boldsymbol{u}}{\p t} = - \nabla p -\boldsymbol{u}\bcdot\bnabla \boldsymbol{u} - \dfrac{2}{E}\hat{\boldsymbol{z}}\times \boldsymbol{u} + \bnabla^{2} \boldsymbol{u} ,
\end{equation}

\begin{equation}\label{eqn:cont}
\bnabla\bcdot\boldsymbol{u}=0 ,
\end{equation}
Here, $\boldsymbol{u}$ represents velocity, $p$ represents an effective pressure that includes centrifugal forces due to the outer boundary (system) rotation. The Ekman number $E = \nu/\Omega_o L^2 = 1/\Omega$, where $\Omega$ is the non-dimensional outer boundary rotation rate. The inner sphere rotation rate (in the rotating frame) can also be similarly non-dimensionalised : $\Delta\Omega = (\Omega_i - \Omega_o)L^2/\nu$. The system and the coordinate system is illustrated in figure \ref{fig:sys}.

The flow problem is then characterised by three non-dimensional numbers - the Ekman number, $E$, the differential rotation $\Delta\Omega/\Omega$, and the radius ratio $\eta = r_i/r_o$, which is set to either $0.33$ or $0.35$. The first is the same as used in H16, while the latter is close to the ratio for Earth's core and has been used in B18 and other previous studies. No-slip boundary conditions allow for the viscous driving of the flow:

\begin{equation}
\begin{array}{ll}
\boldsymbol{u}(r_o) = \boldsymbol{0},\\
\boldsymbol{u}(r_i) = (u_r,u_\theta,u_\phi) = (0,0,\Delta\Omega r_i\sin\theta).
\end{array}
\label{eqn:BC}
\end{equation}
 We numerically solve these equations using two independent pseudo-spectral codes: MagIC \citep{Wicht2002, magic_zenodo} (see \url{https://github.com/magic-sph/magic}) and XSHELLS \citep{xShells} (see \url{https://bitbucket.org/nschaeff/xshells}). The details of the numerical methods can be found in the respective publications. Both codes use the SHTns library \citep{shtns} for spherical harmonic transforms.

\begin{figure}
    \centering
    \includegraphics[width=0.9\tw]{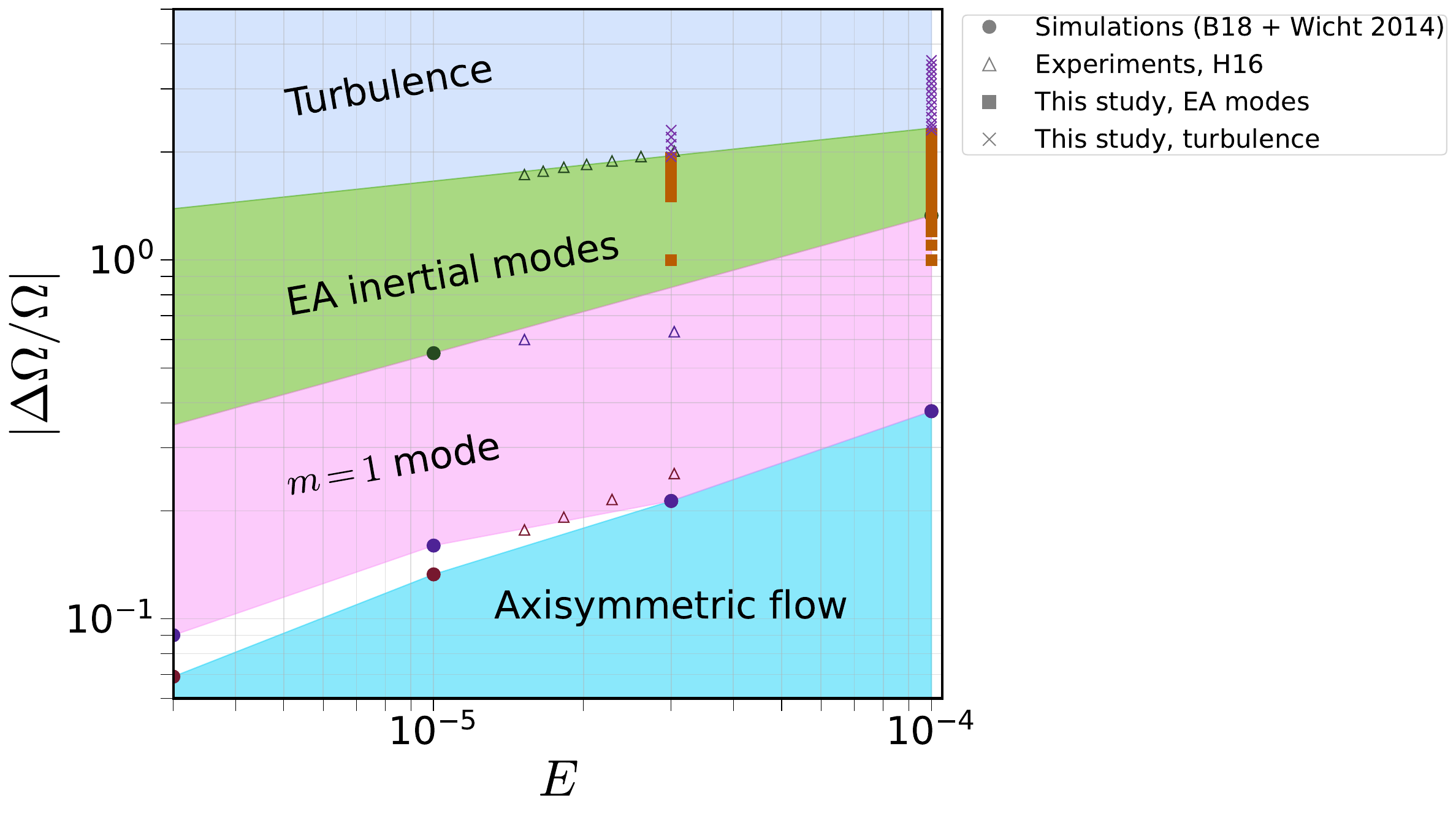}
    \caption{Physical regimes covered in H16, B18, and present study. The horizontal axis shows Ekman number, $E$, while vertical axis shows $|\Ro|$. The open triangles and filled circles represent where various transitions to another regime have been found by experiments of H16 and simulations of B18 and \cite{WichtJFM}, respectively. EA=equatorially antisymmetric. The brown squares and purple crosses show the simulation suites S3 and S4 from table \ref{tab:param}, with squares indicating simulations in the EA inertial mode regime and crosses indicating simulations in the turbulent regime.}
    \label{fig:params}
\end{figure}

As in H16 and B18, we concentrate on the case $\Ro < 0$. The evolution of the flow is studied by keeping the outer boundary rotation (or Ekman number $E$) constant and running a simulation at a fixed $\Ro$ and letting the kinetic energy reach a statistically stationary state. This state is then used as an initial condition to start the simulation for more negative $\Ro$. The various parameters used in simulations and experiments along with the critical $\Ro$ for transition to turbulence are listed in table \ref{tab:param}, with each suite of experiments and simulations  identified by a unique ID (first column). In B18 we already presented the simulation suites S1 and S3. In this study, we have run the rest of the suites, S2, S3a, S4 and S4a, with the suffix `a' representing cases where the parameters are the same as the other case but the simulation is axisymmetric. The case S2 was run to confirm that numerical calculations yield the same critical $\Ro$ for the turbulent transition as the experimental case E1. The ranges of $\Ro$ in the table indicate how the differential rotation was changed within a suit, each using the previous simulation as starting condition  (e.g: $-1.00$ to $-3.50$ means $\Ro$ was made more negative starting from -1). This is important since the behaviour of the system has some amount of hysteresis \citep{EgbersRath95}. Through the rest of the paper, we will mostly focus on simulation suites S3 and S4 with some comparisons with their axisymmetric counterparts S3a and S4a, respectively and with experiments of H16 where appropriate. `Simulations' will thus refer to simulations using MagIC unless otherwise specified. Figure \ref{fig:params} shows a diagram of the different regime transitions identified in simulations (filled circles) and experiments (open triangles). The suites S3 and S4 that is used throughout this paper clearly marked using squares (before transition to turbulence) and crosses (after transition). This does not show the suites S3a and S4a which would largely overlap with S3 and S4.

\subsection{Spectrograms and identification of inertial modes}

It has been shown in previous studies that inertial waves and modes are fundamental instabilities of the spherical Couette system. They obey the linear Euler equation,

\begin{equation}
\dfrac{\p\U}{\p t} = -\nabla p -2\Omega\zh\times\U.
\end{equation}
This can be written as \citep[see e.g.][]{Greenspan, TilgnerTreatise}

\begin{equation}
\dfrac{\p^2}{\p t^2}\nabla^2\U + 4\Omega^2\dfrac{\p^2}{\p z^2}\U = 0,
\end{equation}
which supports plane wave solutions ($\U\propto e^{\ci(\boldsymbol{k}\cdot\boldsymbol{r} - \omega t)}$), called `inertial waves', in an unbounded fluid or bounded global oscillatory modes ($\U\propto e^{\ci(m\phi - \omega t)}$), called `inertial modes', in a bounded container  \citep{Greenspan}.  In both cases, it can be shown that $|\omega|\leq 2\Omega$  where $\omega$ is the frequency associated with a drift in azimuth $\phi$. Here, $\boldsymbol{k}$ and $m$ are the radial wavevector and the azimuthal wavenumber, respectively. For a spherical container, the solutions for inertial modes can be obtained analytically \citep{Bryan1889, KudlickPhD, Greenspan, Zhang2001} and have the form of a spherical harmonic at the surface. Consequently, they are identified using indices $(l,m)$ corresponding to the spherical harmonic degree and order. These, together with the drift frequency $\omega$, uniquely determine a mode. Thus, as in our previous study \citep{Barik2018}, we will denote a mode using the notation $(l,m,\omega/\Omega)$.

\begin{figure}
    \centering
    \includegraphics[width=0.7\tw]{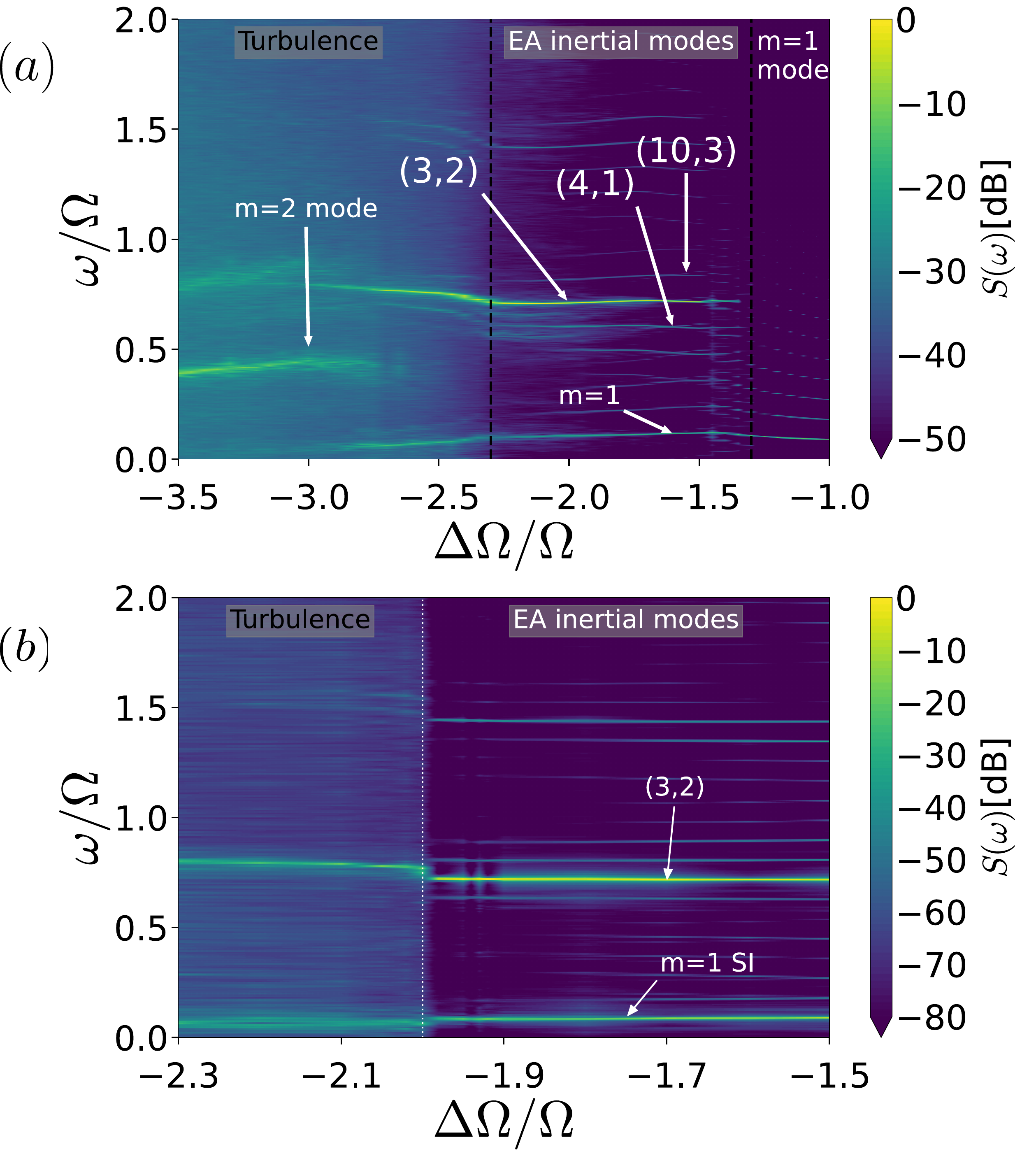}
    \caption{Spectrograms obtained from velocity time series, from simulations. $\omega$ is the angular frequency of the Fourier transform while $S(\omega)$ is the amplitude spectrum. The hydrodynamic regimes are marked and the inertial modes observed have been annotated. SI = Stewartson layer Instability, EA = Equatorially Antisymmetric. 
    (a) Spectrogram from XSHELLS simulations at $E=1.125\times 10^{-4}$. (b) Spectrogram from simulations at $E=3\times 10^{-5}$. }
    \label{fig:spec}
\end{figure}

\begin{figure}
    \centering
    \includegraphics[width=\tw]{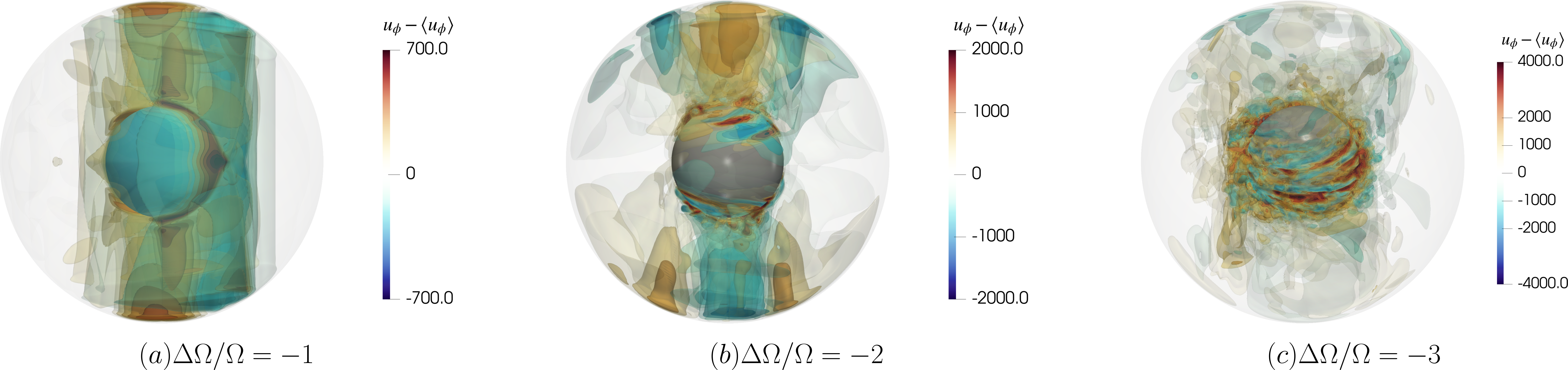}
    \caption{Isosurfaces of non-axisymmetric zonal flow from simulations at $E=10^{-4}$. (a) $\Ro$=-1 is in regime (ii) with the $m=1$ mode, (b) $\Ro=$-2 is in the regime with EA inertial modes and (c) $\Ro=$-3 is in the turbulent regime.}
    \label{fig:flows3d}
\end{figure}

The different hydrodynamic regimes (i) through (v) mentioned in the introduction can be clearly identified using the spectrograms obtained from experimental data. The spectrograms are built by taking the single-sided FFT amplitude spectrum of the radially-averaged azimuthal velocity $u_\phi$ at each $\Ro$. The velocity measurements were performed via Particle Image Velocimetry (PIV) techniques using a laser sheet perpendicular to the axis of rotation. The method is described in further detail in \cite{Hoff2016} and \cite{Hoff2019}. Such spectrograms can also be constructed for simulations where we obtained data for the azimuthal component at eight different locations : $\theta = (\pi/4,3\pi/4)$ and $\phi = (0,\pi/2,\pi,3\pi/2)$, all on a single radial surface, $r = 0.7r_o$, which were stacked after correcting for their phase shift using cross-correlation of the time-series. Thereafter, we performed a Fourier transform of this stacked time series to obtain spectra at each $\Ro$ \citep[see also][]{Barik2018}. The spectrograms obtained from two suites of simulations are shown in figures \ref{fig:spec}(a) and (b), with identified inertial modes denoted by the indices $(l,m)$. Having access to the full three dimensional flow as well as a number of other diagnostics (kinetic energy, spatial spectra etc.) in simulations helps distinguish these different regimes much better. For example, when the first non-axisymmetric $m=1$ mode appears, all the equatorially symmetric $m=1$ spherical harmonic flow coefficients can be seen to oscillate at the same frequency. An analysis of the spectral coefficients, the frequencies in a spectrogram, combined with a visualization of the flow field is used to differentiate between the different regimes.  The non-axisymmetric zonal flow fields in the three different regimes at $E=10^{-4}$ are illustrated in figure \ref{fig:flows3d}. Panel (a) shows the flow at $\Ro=$-1 with the $m=1$ Stewartson layer Instability (SI) clearly visible, panel (b) shows the flow dominated by an equatorially antisymmetric (3,2) inertial mode with some small scale features inside the tangent cylinder, while panel (c) shows the flow in the turbulent regime at $\Ro=-3$, with a lot of small scale features near the inner boundary and a more chaotic flow field.

In the experiments of H16, the inertial modes were identified by comparing their frequencies against frequencies from theoretical works \citep{Zhang2001,WichtJFM}  as well as past experimental works \citep{Kelley2007,Kelley2010,TrianaPhD,Rieutord2012,Matsui2011}. Additional comparisons of morphology of modes were also made against theoretical inertial mode structures in spheres \citep{Zhang2001}. In case of simulations, the inertial modes can be clearly identified by a few different methods. First is by comparing the frequencies observed in the spectrograms to the oscillation frequencies of the spectral spherical harmonic coefficients. This determines the longitudinal symmetry $m$ as well as the equatorial symmetry $(l-m)$ of the mode. The exact mode is then be determined by comparing the frequency to the nearest analytical frequency of inertial modes in a sphere \citep{Zhang2001}, as well as by spectrally filtering out the structure of the mode and comparing it to the theoretical structure.

\section{Identifying transition to turbulence}\label{sec:spectra}

\begin{table}
    \centering
\begin{tabular}{llllll}
ID  & Case      & Ekman number          & $\eta = r_i/r_o$ & $\Ro$ range        & $\Ro_c$ \\ \hline
E1  & BTU C-S   & $3.043\times 10^{-5}$ & 0.33             & $-2.37$ to $-0.19$ & $-2.011$                             \\
E2  & BTU C-S   & $1.522\times 10^{-5}$ & 0.33             & $-2.36$ to $-0.2$  & $-1.730$                             \\
S1  & XSHELLS   & $1.125\times 10^{-4}$ & 0.33             & $0.00$ to $-3.50$  & $-2.3$                               \\
S2  & MagIC     & $3.043\times 10^{-5}$ & 0.33             & $-2.10$ to $-2.00$ & $-2.015$                             \\
S3  & MagIC     & $10^{-4}$             & 0.35             & $-1.00$ to $-3.50$ & $-2.3$                               \\
S3a & MagIC     & $10^{-4}$             & 0.35             & $-0.01$ to $-3.50$ & $-2.1$                               \\
S4  & MagIC     & $3\times 10^{-5}$     & 0.35             & $-0.18$ to $-2.30$ & $-2.0$                              \\
S4a & MagIC     & $3\times 10^{-5}$     & 0.35             & $-0.01$ to $-2.30$ & $-1.9$                              
\end{tabular}
    \caption{Parameters for different experiments at BTU C-S and simulations using MagIC and XSHELLS used in this study. Label `a' denotes axisymmetric simulations. $\Ro_c$ indicates the critical $\Ro$ for transition to turbulence. }
    \label{tab:param}
\end{table}

\begin{figure}
    \centering
    \includegraphics[width=0.7\textwidth]{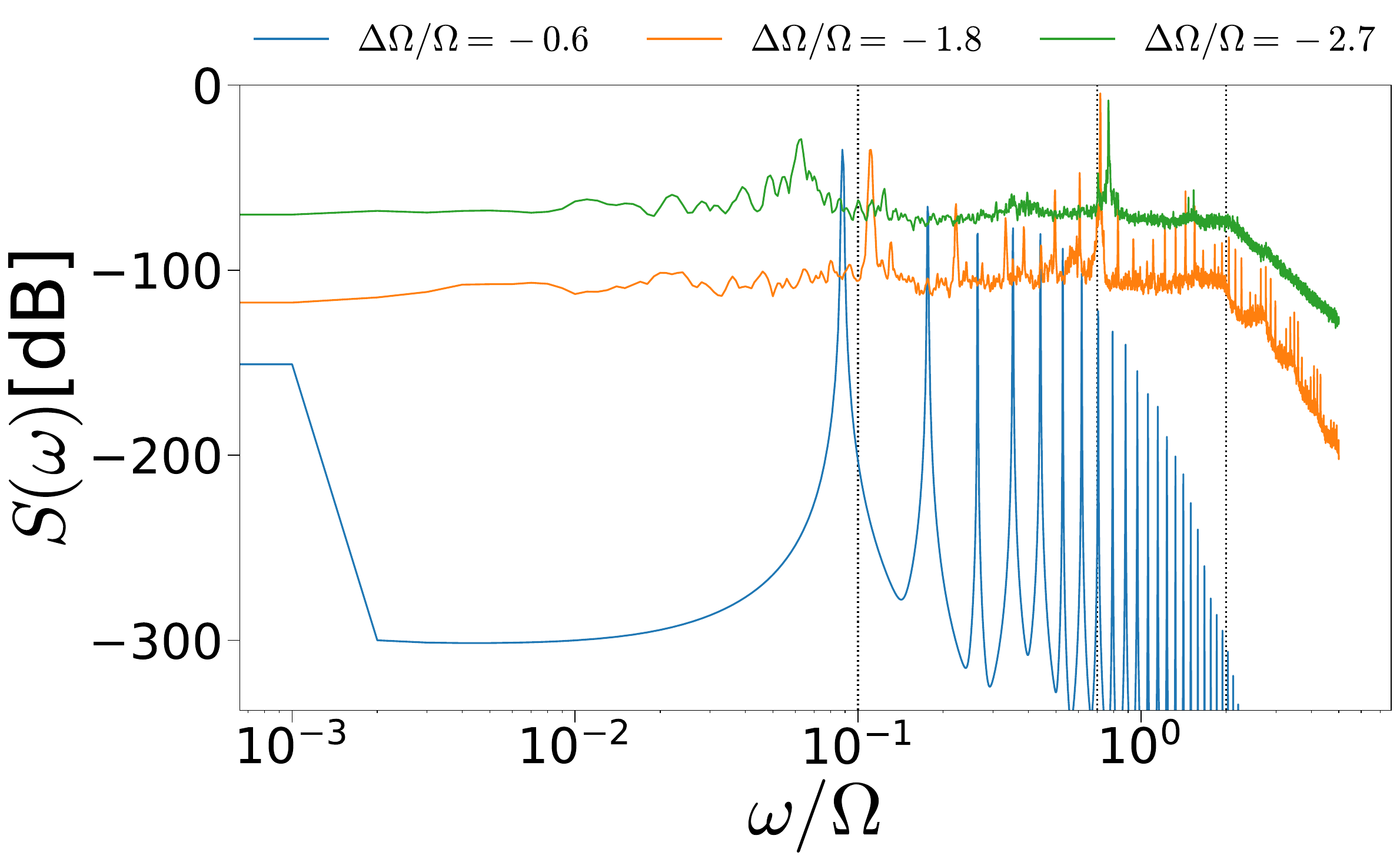}
    \caption{Temporal spectra at three different values of differential rotation, each in a different hydrodynamic regime. The horizontal axis shows the Fourier frequency $\omega$ scaled with the outer boundary rotation rate $\Omega$ while the vertical axis shows the amplitude spectrum $S(\omega)$. Vertical dotted lines mark $\omega/\Omega = 0.1, 0.7, 2$.}
    \label{fig:specRoRatio}
\end{figure}

In experiments as well as simulations, the temporal spectra help us determine the transition to turbulence. We examine here the spectra at individual $\Ro$ values from the XSHELLS spectrogram presented in figure \ref{fig:spec}(a). We have selected three representative $\Ro$ values, $\Ro = (-0.6, -1.8, -2.7)$, which lie in regimes (iii), (iv) and (v), respectively, as shown in figure \ref{fig:specRoRatio}. At $\Ro = -0.6$, the spectrum consists of only discrete peaks at the drift frequency of the $m=1$ SI and its higher multiples. In the EA inertial mode regime, at $\Ro = -1.8$ (orange), there is a drastic change in the nature of the spectrum and it consists of a nearly flat background for $\omega/\Omega \leq 2$ and a sharp decay for larger Fourier frequencies. The frequencies of the $m=1$ SI (around $\omega/\Omega$ = 0.1) and of the dominant inertial mode ( $(3,2)$ mode, around $\omega/\Omega = 0.7$) are the most clearly visible peaks on top of the flat background. A flat background of energy for $0<\omega<2\Omega$ and a subsequent decay demonstrates the fact that most of the kinetic energy in the flow manifests in inertial waves and is characteristic of inertial wave turbulence \citep[e.g.][]{Duran2013, Clark2015}.  Thus, there is some amount of inertial wave turbulence already present in the EA inertial modes regime. This can be seen in the small scales visible inside the tangent cylinder in figure \ref{fig:flows3d} (b), close to the inner boundary. However, the large scale inertial mode still carries the dominant amount of energy in this regime.

What we define as the `turbulent' regime in this study is characterised by a further sudden increase in this flat background spectrum of inertial waves, as seen for $\Ro = -2.7$, while the decay beyond $\omega/\Omega = 2$ becomes less steep. Consequently, the peaks for the $m=1$ SI and the $(3,2,0.666)$ mode, despite having similar energies as for $\Ro = -1.8$, are now less prominent with respect to the background. The small scale inertial wave turbulence is no longer limited to inside the tangent cylinder, but now can be seen in the bulk as well (figure \ref{fig:flows3d} (c)) and thus, the global large scale inertial mode no longer carries a huge fraction of the energy. The typical decay of the spectrum beyond $2\Omega$ has also been observed by H16 \citep[figure 10 of][]{Hoff2016} and in the 3-meter experiment \cite[figures 6.3 and 6.15 of][]{TrianaPhD}. The shallower decay of the spectrum beyond $2\Omega$ in the turbulent regime shows a decrease in the influence of rotation which results in a greater content of energy for $\omega > 2\Omega$. This is consistent with the fact that smaller scales and increased flow velocities in the turbulent regime lead to a dominance of advection over the effect of the Coriolis force. The change in the force balance is further discussed in section \ref{sec:ForceBalance}.

\begin{figure}
    \centering
    \includegraphics[width=\tw]{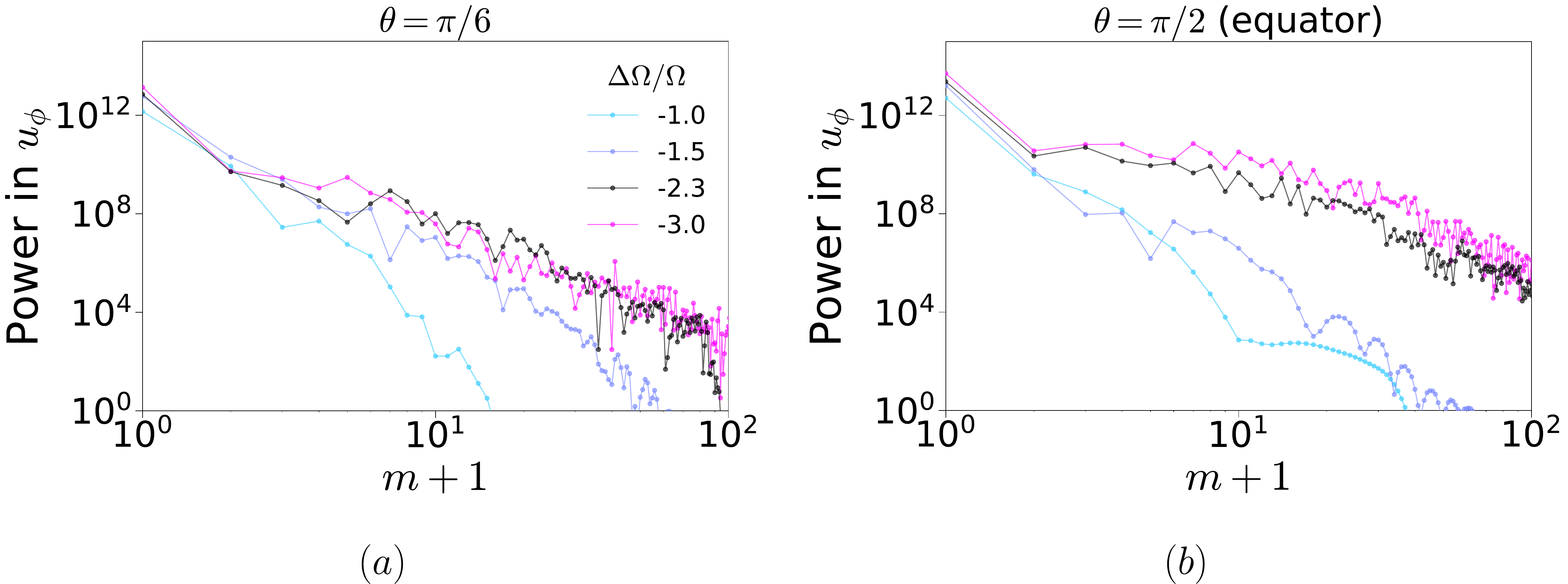}
    \caption{Change in the power spectrum of the the zonal flow $u_\phi$ with $\Ro$ at different latitudes at a fixed radius $r/r_o = 0.354$. Horizontal axis shows the azimuthal wavenumber $m$ while vertical axis shows the power in a single wavenumber. $\Ro=-2.3$ which marks the transition to turbulence is plotted with a black line.}
    \label{fig:up_m}
\end{figure}

Pseudo-spectral codes provide direct information on different flow length scales and hence spatial spectra of kinetic energy. Figure \ref{fig:up_m} shows the change in energy spectrum in the zonal flow with $\Ro$ at different colatitudes, very close to the inner boundary at $r/r_o = 0.354$. In panel (a), we see that for all $|\Ro| \geq 1.5$ in the EA inertial mode regime, there already is a significant amount of energy in the smaller scales (high $m$) inside the tangent cylinder. In panel (b), we find that the energy in the smaller scales are high for $|\Ro| \geq 2.3$, indicating that the boundary layer at the inner boundary gets progressively destabilised at lower latitudes as $\Ro$ becomes more negative. The turbulent regime sets in at $\Ro = -2.3$ and its significance is that the boundary layer at the equator gets destabilised.

\begin{figure}
    \centering
    \includegraphics[width=0.7\tw]{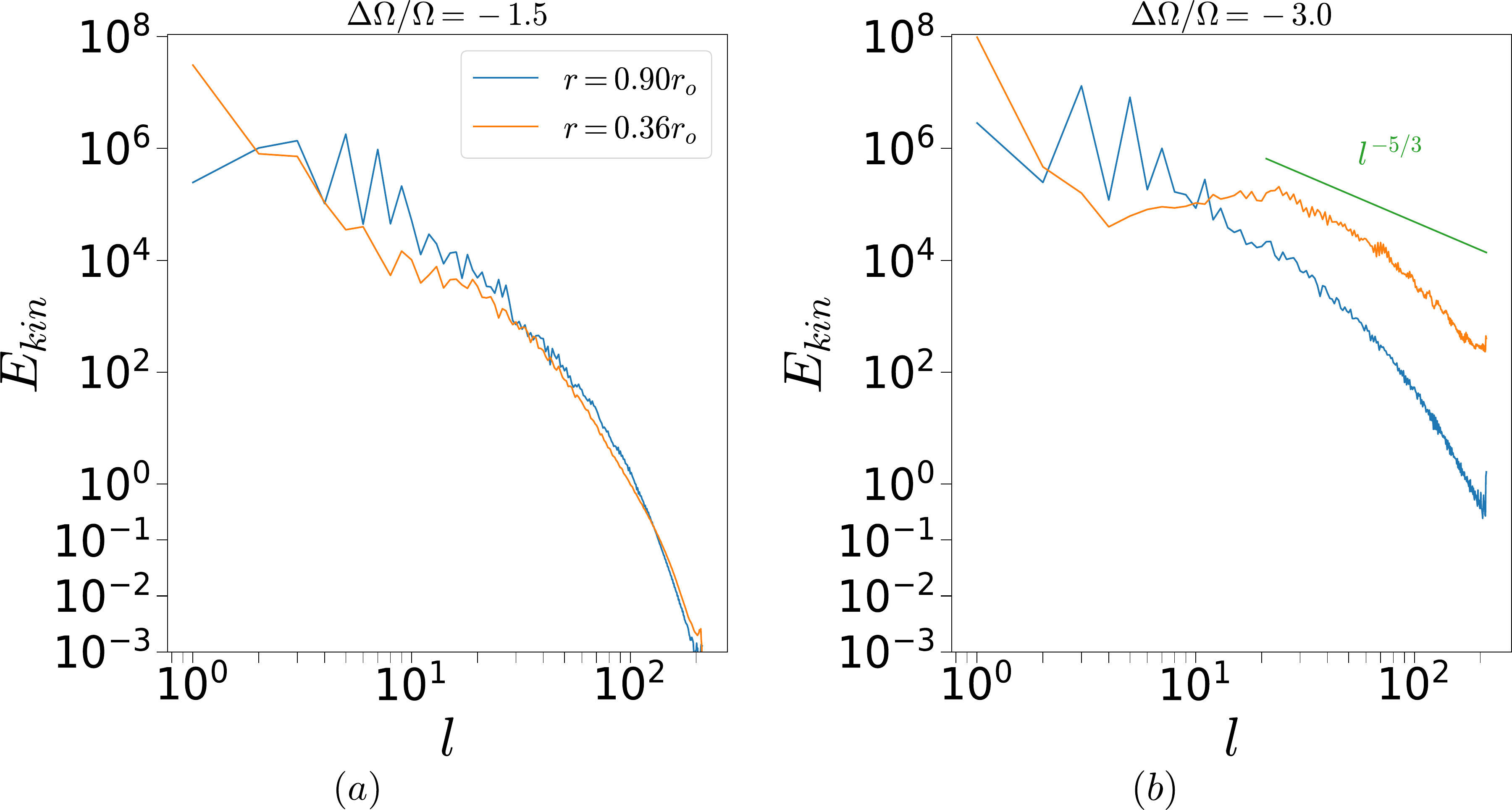}
    \caption{Kinetic energy spectra from simulations at $E=10^{-4}$, shown at different radial levels. (a) shows the case for $\Ro = -1.5$, before the transition to turbulence, while (b) shows the case for $\Ro = -3$, after the transition.}
    \label{fig:2D_spat_spec}
\end{figure}

Figure \ref{fig:2D_spat_spec} shows the total kinetic energy spectra with respect to spherical harmonic order $l$ at different radial levels from S3 simulations at $E=10^{-4}$. The onset of turbulence in the spatial spectra is characterised by an increase in energy in the small scales in general and close to the inner boundary in particular, which is the region of the highest flow speeds and thus most extreme Reynolds numbers. The system is driven by imposed differential rotation at the largest system scale. The energy then cascades to smaller scales via the different instabilities and non-linear interactions. This cascade becomes decisively more efficient in the turbulent regime. The decrease in the influence of rotation can be seen in the spatial spectrum at large spherical harmonic degree as it gets progressively closer to a classic Kolmogorov -5/3 spectrum, as shown in figure \ref{fig:2D_spat_spec} (b).

\section{Flow analysis}\label{sec:flowAnalys}

\subsection{Mean flow and angular momentum transport}\label{subsec:MeanFlow}

\begin{figure}
    \centering
    \includegraphics[width=\textwidth]{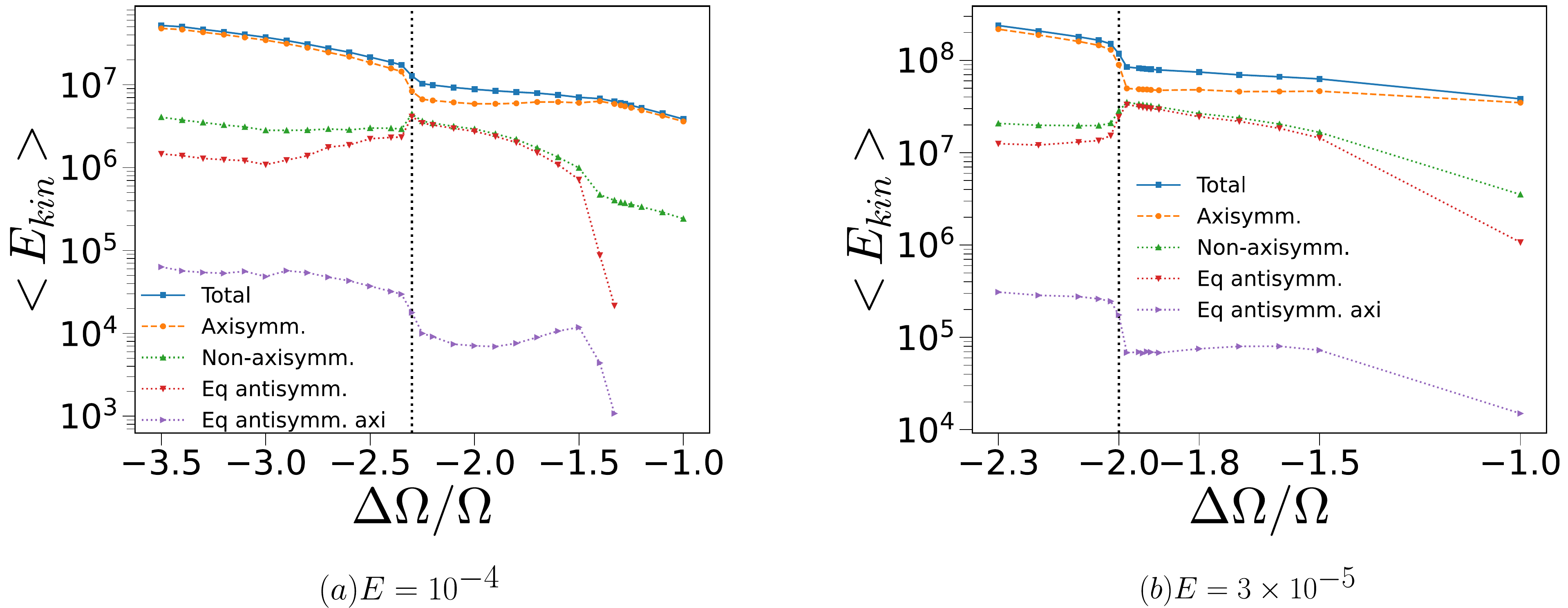}
    \caption{Change in kinetic energy with differential rotation. As the flow transitions to turbulence (marked with vertical dotted lines), there is a sudden `burst' in axisymmetric, mostly zonal, kinetic energy. The non-axisymmetric flow contributions, on the other hand, decrease in amplitude.}
    \label{fig:KE}
\end{figure}

\begin{figure}
    \centering
    \includegraphics[width=\textwidth]{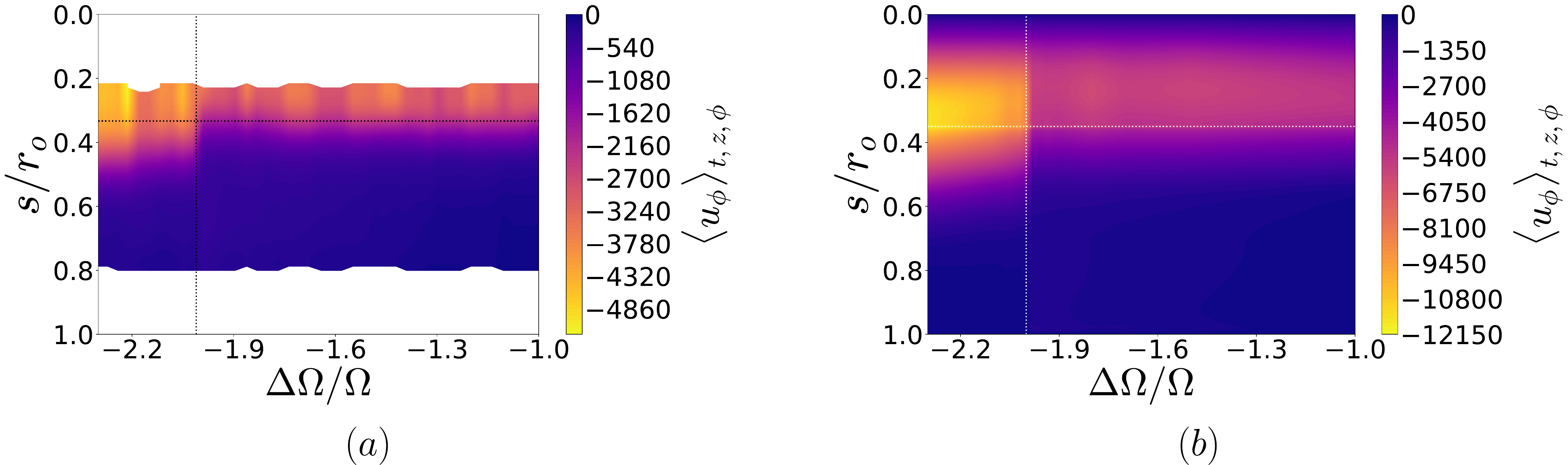}
    \caption{$z,\phi$ and time-averaged zonal flow velocity. The horizontal axis shows differential rotation while the vertical axis shows the cylindrical radial coordinate, scaled to the outer boundary $s/r_o$. The horizontal dotted line marks the tangent cylinder while the vertical dotted line marks the critical differential rotation for the transition to turbulence. (a) shows zonal flow from the experiments of H16 at $3.043\times 10^{-5}$ while (b) shows the same from simulations at $E=3\times 10^{-5}$.}
    \label{fig:VphiMean}
\end{figure}

\begin{figure}
    \centering
    \includegraphics[width=0.7\tw]{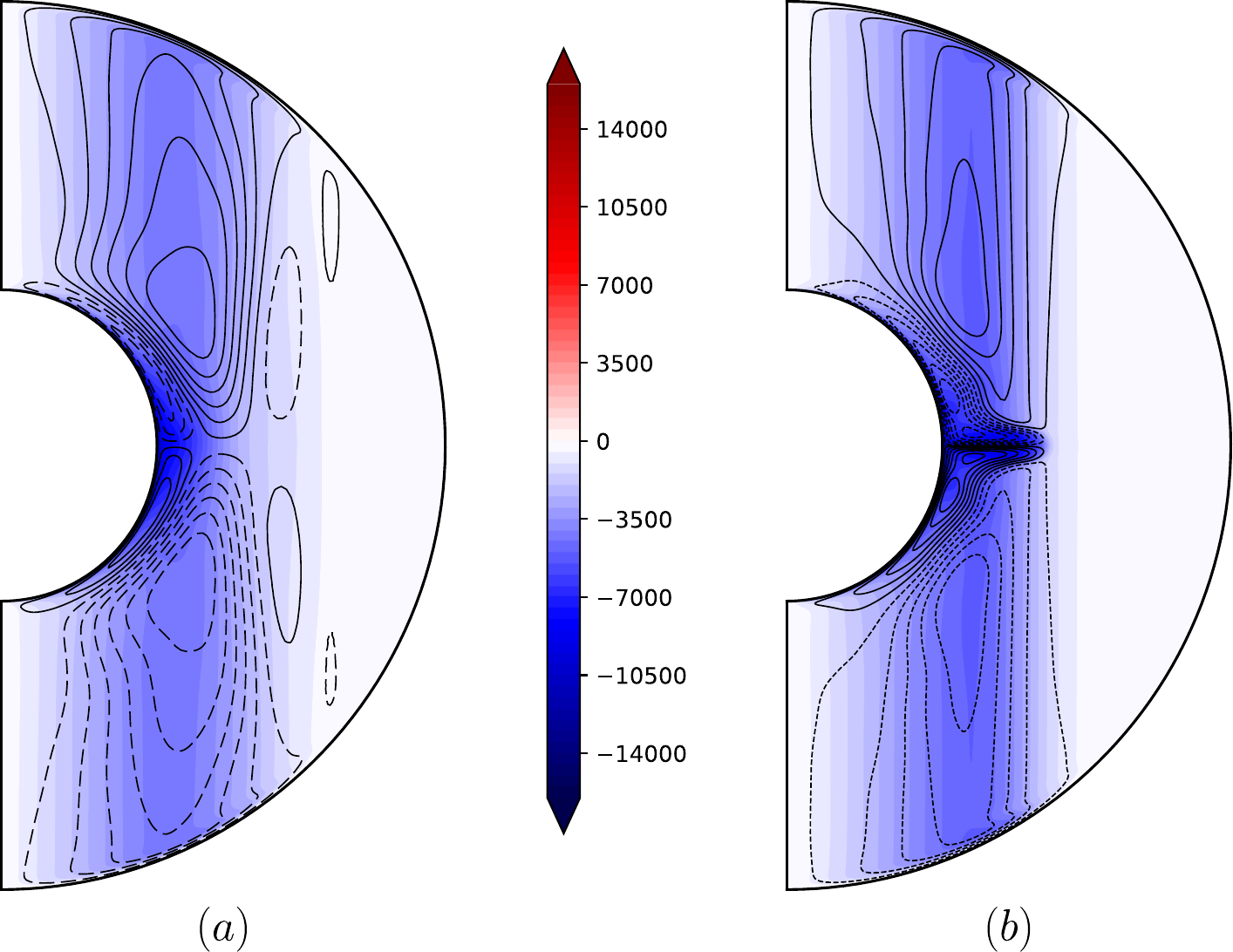}
    \caption{Zonal flow and streamlines of meridional circulation from simulations at $E=10^{-4}$. Dashed (solid) lines represent counterclockwise (clockwise) circulation. (a) shows the case for a turbulent 3D simulation at $\Ro = -3.5$ while (b) shows the same for an axisymmetric simulation at $\Ro = -3.5$. Both are averaged in time and azimuth. Colours indicate zonal flow with blue being retrograde and red, if any, being prograde.}
    \label{fig:merLines}
\end{figure}

The transition to turbulence is characterised by a sudden increase in the axisymmetric flow, while there is a drop in the non-axisymmetric kinetic energy (figure \ref{fig:KE}). The axisymmetric flow is clearly dominated by the zonal component, which is by a factor $E^{-1/2}$ larger than the meridional circulation. Both components increase upon turbulence onset. The non-axisymmetric component is dominated by the equatorially antisymmetric part owing to the presence of the EA inertial modes before the transition to turbulence. This changes upon turbulence onset when EA inertial modes lose their energy. 
Figure \ref{fig:VphiMean}(a) and (b) illustrate that the mean zonal flow not only intensifies but also starts to spread beyond the TC. The panels show the mean zonal flow, averaged in the $z$- and $\phi$-directions and in time, as a function of the cylindrical radial distance $s$ and the differential rotation rate for experiments E1 (a) and simulations S4 (b). 
Before the transition to turbulence (vertical lines), the zonal flow roughly resembles the Proudman solution for spherical Couette flow\citep{Proudman56}, staying restricted to the region inside the TC (horizontal lines). Beyond the transition, the zonal flow is significantly more vigorous and extends beyond the TC. The mean flow behaves the same way for simulations at $E=10^{-4}$.
From table \ref{tab:param}, we can compare the onset of turbulence for full 3D simulations S3 and S4 and their axisymmetric counterparts, S3a and S4a.
For $E=10^{-4}$, turbulence sets in at a $13$\% lower differential rotation rate, at $E=3\times 10^{-5}$ the difference has reduced to $5$\%.  
Figure \ref{fig:merLines} compares the time and azimuthally averaged zonal flow and medirional circulation in the turbulent regime at $E=10^{-4}$ of a 3D simulation (a) and an axisymmetric one (b). Both cases show an additional pair of rolls along the inner boundary. These rolls represent an outward flow at the inner boundary which can contribute to advection towards the equator and then into the region outside the tangent cylinder. In the axisymmetric case (panel (b)), the inner roll pair is more pronounced than in the 3D case at the same parameters (panel a). In addition, an additional pair of rolls develops near the inner boundary equator and subsequently joins the set of rolls at higher latitudes to form a continuous radial jet. This jet-like feature is not seen in the 3D case. The overall structure of the zonal flows is very similar in the two cases, axisymmetric turbulence is also characterized by a spreading of zonal flows beyond the TC.

One can notice that the meridional circulation in panel (a) looks equatorially asymmetric compared to panel (b). Beyond the onset of the EA inertial modes, the 3D simulations continue to possess an equatorially antisymmetric component of kinetic energy (figure \ref{fig:KE}), while the kinetic energy for the axisymmetric simulations continues to remain purely equatorially symmetric, even beyond the onset of turbulence. Thus, the symmetry breaking with respect to the equator seems unique to the presence of non-axisymmetric flows. The extension of the mean flow into the bulk along with the additional roll pair seems to push the Stewartson layer away from the TC. Whether we should still call this a Stewartson layer is unclear.  As already discussed by \cite{WichtJFM}, the appearance of a new pair of rolls close to the inner boundary indicates that the Coriolis force due to the outer boundary rotation ceases to be dominant. We expect this to happen when $\Ro$ becomes smaller than $-1$. At $\Ro=-1$, the inner boundary is at rest in the inertial frame. For $\Ro \leq -1$ it rotates in the opposite direction to the outer boundary. When $\Ro$ is negative enough, centrifugal forces drive an outward flow at the inner boundary that gives rise to the additional meridional roll pair. The transition from Coriolis-force dominated dynamics to inertia dominated dynamics should start close to the inner boundary where the effective rotation in the inertial frame is minimal.

\begin{figure}
    \centering
    \includegraphics[width=0.6\textwidth]{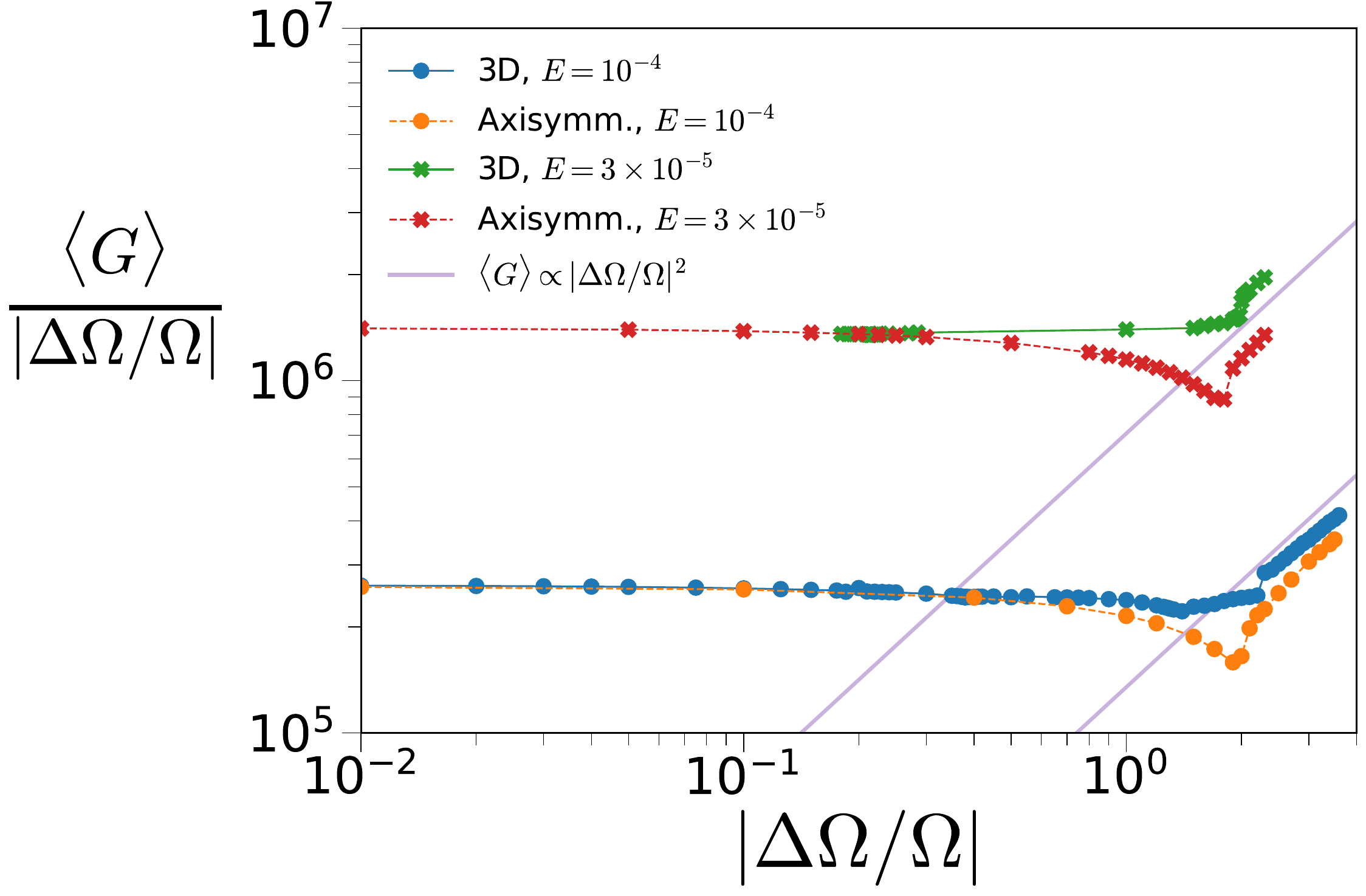}
    \caption{Torque applied to the inner sphere and its variation with differential rotation magnitude, compensated by the linear scaling. The solid purple lines show a quadratic scaling.}
    \label{fig:torqScaling}
\end{figure}

Turbulence creates small scale flow and transports angular momentum more efficiently from the inner boundary to the bulk of the flow outside the tangent cylinder. This increases the viscous friction at the inner boundary so that a larger torque at the inner boundary is required to maintain the flow. Figure \ref{fig:torqScaling} shows the increase of the viscous torque on the inner core with $\Ro$ for simulations at two different Ekman numbers. Before the onset of turbulence, the torque is simply proportional to $|\Ro|$ and scales like $G\sim |\Ro|^\alpha$ with $\alpha=1$, as shown with the compensated plot. In the turbulent regime, however, the torque increases more steeply with $\alpha \sim 2$. \cite{DanZ} reported approaching $\alpha=2$ in the 3-metre experiment in Maryland for a non-rotating outer boundary. The torque scalings for the axisymmetric simulations show a similar behaviour but the torque becomes smaller than in the 3D simulations where non-axisymmetric instabilities provide a more efficient transport of angular momentum. In conclusion, the instability responsible for the onset of turbulence is predominantly an instability of the axisymmetric flow. The weaker non-axisymmetric flow components help stabilise the flow and yield a later onset.

\subsection{Inertial modes}

\begin{figure}
    \centering
    \includegraphics[width=\textwidth]{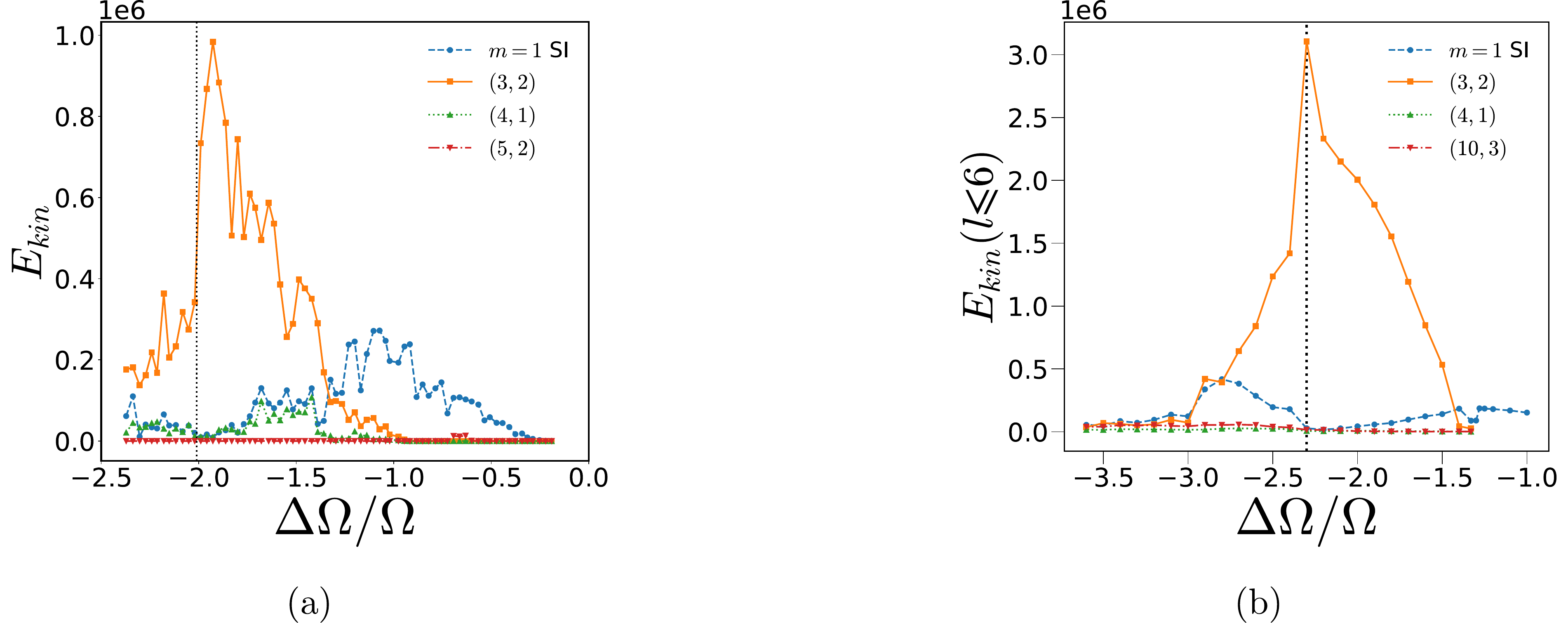}
    \caption{Kinetic energy of the different inertial modes. (a) shows the data from experiments E1 at $E=3.043\times 10^{-5}$ while (b) shows data from simulations S3 at $E=10^{-4}$. Both show the same features of the dominant inertial mode $(3,2)$ having a sudden drop in its kinetic energy. Vertical dotted lines mark the transition to turbulence in each case.}
    \label{fig:modEner}
\end{figure}

The large scale EA inertial modes that get excited in regime (iv) continue to exist after the transition to turbulence. There is, however, a jump in the inertial mode frequencies. This can clearly be seen in the `brightest' spectral lines in both panels of figure \ref{fig:spec}. This goes together with the sudden spreading of the background zonal flow beyond the TC causing further deformation of the inertial modes, as shown in B18. In both 3D simulation suites that we studied, the flow is dominated by the inertial mode $(3,2,0.666)$ when turbulence sets in. After the transition, the mode loses at least half of its energy but still clearly dominates against the broadband turbulent background, as shown in figure \ref{fig:modEner} for experiments E1 and simulations S3. The energy estimates were determined by using a frequency filter on velocity obtained in experiments. In case of simulations, the energy in the large scale spherical harmonic coefficients (order $l \leq 6$) of the equatorial and azimuthal symmetry corresponding to a mode was used to estimate the energy in a mode.

\begin{figure}
    \centering
    \includegraphics[width=\tw]{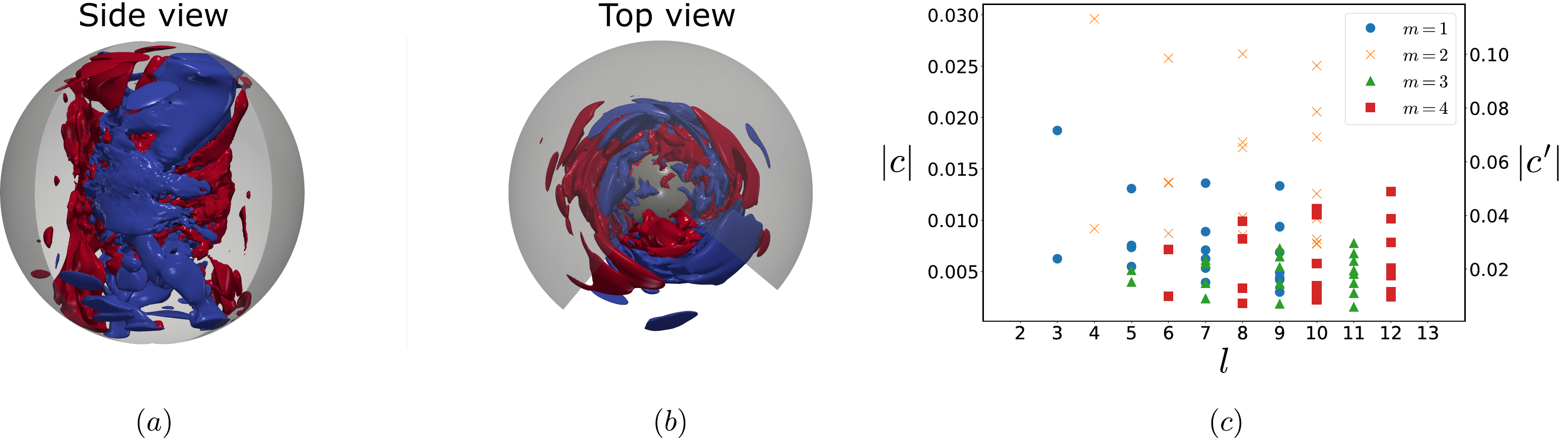}
    \caption{A new $m=2$ mode emerging in the turbulent regime. (a) shows the 3D side view of a snapshot from a MagIC simulation at $E=10^{-4}, \Ro = -3$. The isosurfaces show non-axisymmetric zonal flow with red (blue) being positive (negative) at $u_\phi = \pm 500$. (b) shows the top view of the same. (c) shows the projection of the flow onto equatorially symmetric inertial modes.}
    \label{fig:vpRo-3}
\end{figure}

In both the numerical simulations S3 at $E=10^{-4}$ (MagIC) and S1 at $E=1.125\times 10^{-4}$ (XSHELLS), a new $m=2$ mode emerges around $\Ro = -2.9$ with a frequency of $\omega/\Omega \approx 0.4$. The mode is visualised at $\Ro = -3$ in figure \ref{fig:vpRo-3}(a). We project snapshots of the flow velocity $\U$ and its non-axisymmetric part at different times onto equatorially symmetric inertial modes of a sphere $\Q_j e^{i(m\phi - \omega_j t)}$, similar to B18,
\begin{equation}
\begin{split}
    \U &= \sum c_j \Q_j e^{i(m\phi - \omega_j t)}\\
    \U - \langle\U\rangle_\phi &= \sum c_j' \Q_j e^{i(m\phi - \omega_j t)}
\end{split}
\end{equation}

The projection coefficients are normalised by $\left[\int \U\cdot\U dV \int \Q_j\cdot\Q_j^\dagger dV \right]^{1/2}$ (or $\left[\int \left(\U - \langle\U\rangle_\phi\right)\cdot\left(\U - \langle\U\rangle_\phi\right) \int \Q_j\cdot\Q_j^\dagger dV \right]^{1/2}$ in case of $c_j'$). The corresponding projection coefficients $c$ and $c'$, respectively are shown in panel (b). It is clear that a single inertial mode cannot be used to characterise this flow structure, with dominant contributions from all modes with $m=2, l \leq 10$ that were analysed. We also could not find other modes that form triadic resonances with this mode. 

\section{Force balance} \label{sec:ForceBalance}

\begin{figure}
    \centering
    \includegraphics[width=0.8\tw]{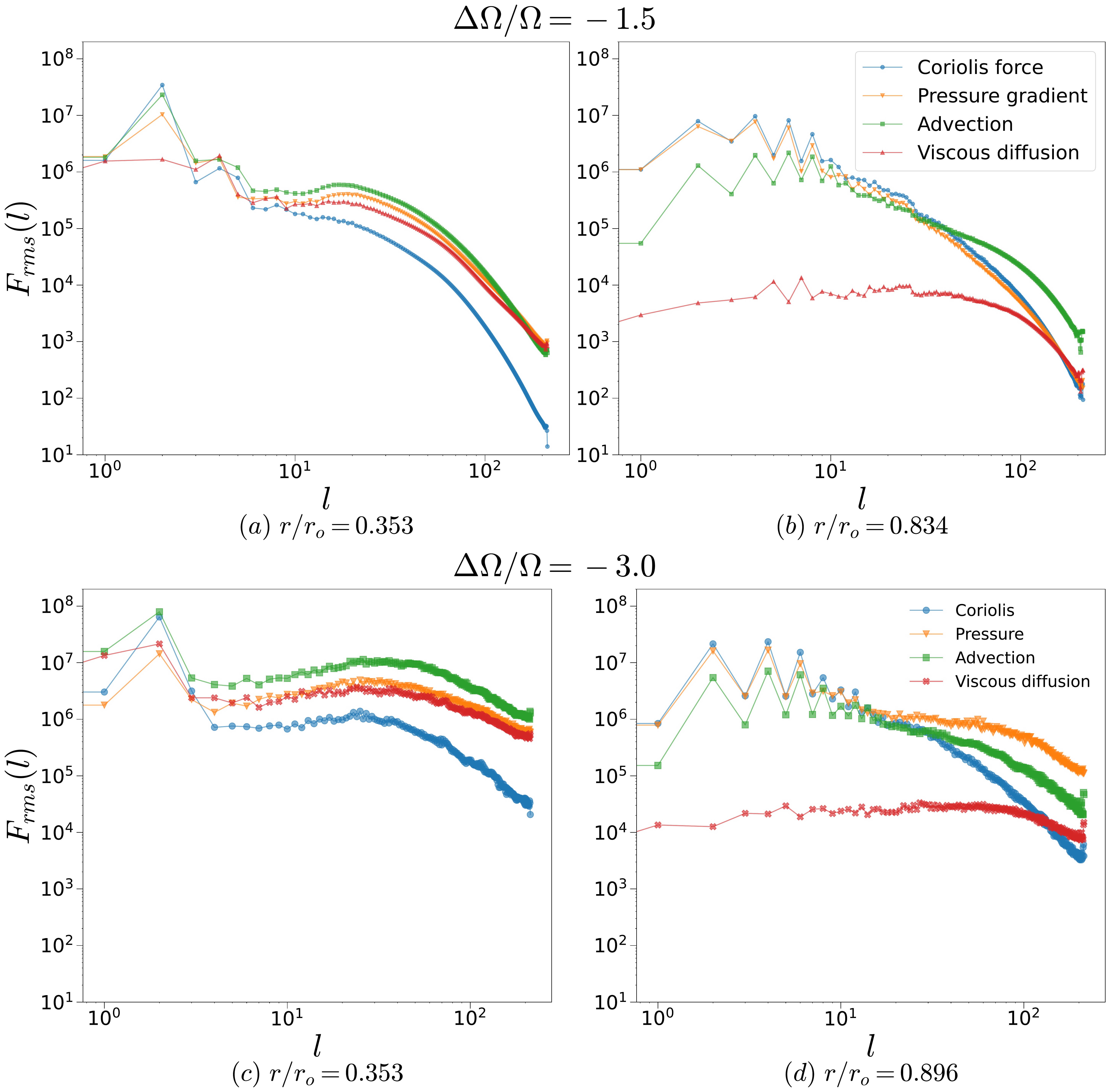}
    \caption{RMS spectra of different forces in the Navier-Stokes equation at two different radial levels, from simulations at $E=10^{-4}$. Horizontal axes show spherical harmonic degree while vertical axes show RMS values. (a) and (b) show the case at $\Ro = -1.5$ in the laminar regime while (c) and (d) show it for $\Ro = -3$ in the turbulent regime. All plots have the same scale on the vertical axis.}
    \label{fig:ForceBal2D}
\end{figure}

The transition to the turbulent regime for both S3 and S4 goes along with a sudden rise in the nonlinear term ($\U\cdot\nabla\U$). As a consequence, advection rather than Coriolis becomes the dominant force. To understand the force balance at different length scales, we decompose the magnitude of each force $F$ into spherical harmonics,

\begin{equation}
    F(r) = \sum_{l=0}^{l_{max}}\sum_{m=0}^l F_{lm}(r) Y_{lm}(\theta,\phi)~,
\end{equation}
where, $Y_{lm}(\theta,\phi)$ denotes a spherical harmonic of degree $l$ and order $m$. We then investigate the magnitude of forces, at different specific spherical harmonics degrees $l$ and radius levels, similar to \cite{Schwaiger2019},

\begin{equation}
    F_{rms}^2(l,r) =  \dfrac{1}{V} \sum_{m=0}^{l} r^2 |F_{lm}(r)|^2~,
\end{equation}
where $V = 4/3 \pi (r_o^3 - r_i^3)$ is the volume of the spherical shell. Figure \ref{fig:ForceBal2D} compares the respective spectra for two simulations at $E=10^{-4}$, one before the transition to turbulence (panels (a) and (b) ) and one after (panels (c) and (d)). This is done for two different radial levels, one near the inner boundary and one in the bulk. At large scales (low $l$), the leading order force balance near the inner boundary is dominated by advection and the Coriolis force while the dynamics in the bulk is determined by a geostrophic balance between the Coriolis force and the pressure gradient. This remains true for both before as well as after the transition to turbulence. At small scales (large $l$), after the transition to turbulence, there is a clear dominance of advection close to the inner boundary. This leads to the large scale flow in the system to be aligned with the rotation axis even in the turbulent regime, while small scale flows dominate close to the inner boundary. This can be seen in the 3D flow visualisation in figure \ref{fig:flows3d}(c) combined with the zonal flow visualised in \ref{fig:merLines}.

\begin{figure}
    \centering
    \includegraphics[width=\tw]{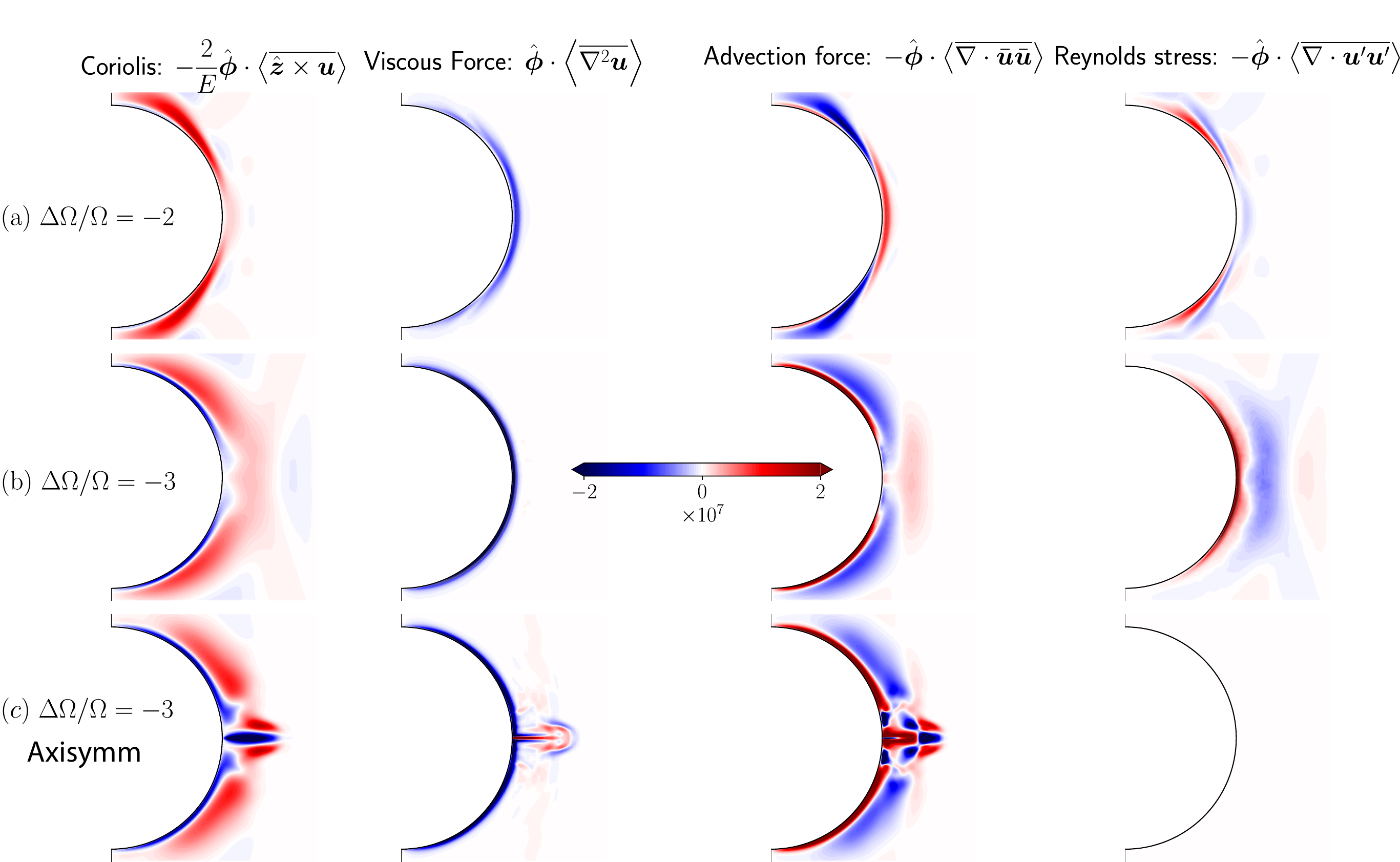}
    \caption{Different terms of the RANS eqution \ref{eqn:reynolds} computed near the inner boundary for simulations at $E=10^{-4}$. (a) shows the case for $\Ro = -2$, in the inertial mode regime, and (b) for $\Ro = -3$, in the turbulent regime. (c) shows a turbulent axisymmetric simulation at $\Ro = -3$. The plots are clipped at a cylindrical radius of $s/r_o = 0.65$ and a vertical extent of -0.6 to 0.6.}
    \label{fig:reynolds}
\end{figure}

We investigate the effect of the turbulent small scales on angular momentum transport using the azimuthal component of the Navier-Stokes equation. Separating the flow velocity and pressure into mean and fluctuating parts and a subsequent mean in azimuth and time gives us the Reynolds averaged Navier-Stokes (RANS) equation for the mean zonal flow:

\begin{equation}\label{eqn:reynolds}
\begin{split}
    -\dfrac{2}{E} \hat{\boldsymbol{\phi}}\cdot\left\langle{\overline{\hat{\boldsymbol{z}}\times\U}}\right\rangle + \hat{\boldsymbol{\phi}}\cdot\left\langle\overline{\nabla^2\U}\right\rangle - \hat{\boldsymbol{\phi}}\cdot\left\langle\overline{\nabla\cdot\bar{\U}\bar{\U}}\right\rangle - \hat{\boldsymbol{\phi}}\cdot\left\langle\overline{\nabla\cdot\U'\U'}\right\rangle = 0,
\end{split}    
\end{equation}
where a bar denotes a mean in azimuth, $\langle\rangle$ denotes an average in time and prime denotes a non-axisymmetric part. We use about 700 snapshots of the 3D flow at $\Ro = -2$ in the inertial mode regime and about 1000 snapshots at $\Ro = -3$ in the turbulent regime at $E=10^{-4}$, to compute the terms above, corresponding to 100 rotations of the outer boundary in each case. In addition, we use a time-averaged flow file for computing the terms for an axisymmetric case at $\Ro = -3$, where the Reynolds stress $\left\langle\overline{\nabla\cdot\U'\U'}\right\rangle$ is absent. The results are shown in figure \ref{fig:reynolds}. In all cases, as expected, viscous forces are a dominant contributor to the zonal flow acceleration near the inner boundary. In the inertial mode regime (figure \ref{fig:reynolds} (a)), there is very little zonal flow generation and hence, very little forcing outside the tangent cylinder. Here the advection force $\left\langle\overline{\nabla\cdot\bar{\U}\bar{\U}}\right\rangle$ balances the viscous force near the equator and the Coriolis force away from the equator. Beyond the transition to turbulence (panel (b)), Reynolds stresses near the inner boundary balance the viscous force in this region, while the advective force provides the balance away from the equator. Slightly away from the boundary, the advective force balances the Coriolis force. In the axisymmetric turbulent case (panel (c)), in the absence of Reynolds stresses, the advective force balances both the Coriolis force as well as the viscous drag.

In the case of 3D turbulence, the small scales in the bulk of the fluid lead to Reynolds stresses that play the dominant role in forcing the zonal flow outside the TC, while in the axisymmetric case, the advection due to the strong radial jet plays the same role. The resultant efficient transport of angular momentum manifests itself in a change in torque scaling. Before the transition to turbulence, the zonal flow is restricted to the TC and its amplitude is linearly dependent on $\Delta\Omega$, just as shown by \cite{Proudman56}. Thus the torque on the inner sphere, $G = r_i \int \tau_{r\phi} dS = r_i\int \p/\p r(u_\phi/r) dS$, where $dS = r_i\sin\theta d\theta d\phi$ is the differential surface area at the inner boundary, is also proportional to $\Delta\Omega$. Beyond the transition to turbulence, the Reynolds stresses and the advective force significantly contribute to the zonal flow near the equator and become the major player in enhancing angular momentum transport. Their quadratic nature thus explains the quadratic scaling law in the turbulent regime.

\section{Instability near the inner boundary}\label{sec:BL}

\begin{figure}
    \centering
    \includegraphics[width=\textwidth]{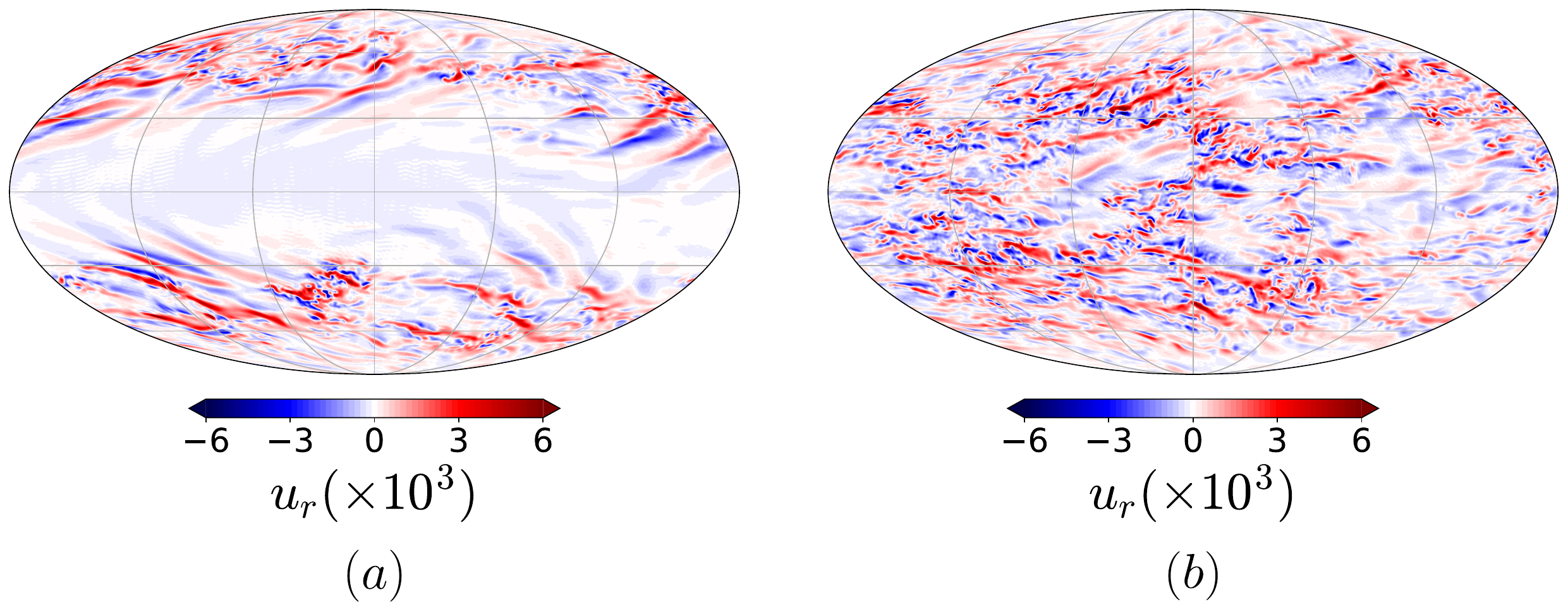}
    \caption{Mollweide projection of radial velocity near the inner boundary at $r/r_o = 0.36$. Both simulations are for the suite S4 at $E=3\times 10^{-5}$. (a) shows the case for $\Ro = -1.98$ before the transition to turbulence while (b) shows the case for $\Ro = -2$, after the transition to turbulence.}
    \label{fig:vr3e-5}
\end{figure}

\begin{figure}
    \centering
    \includegraphics[width=0.7\textwidth]{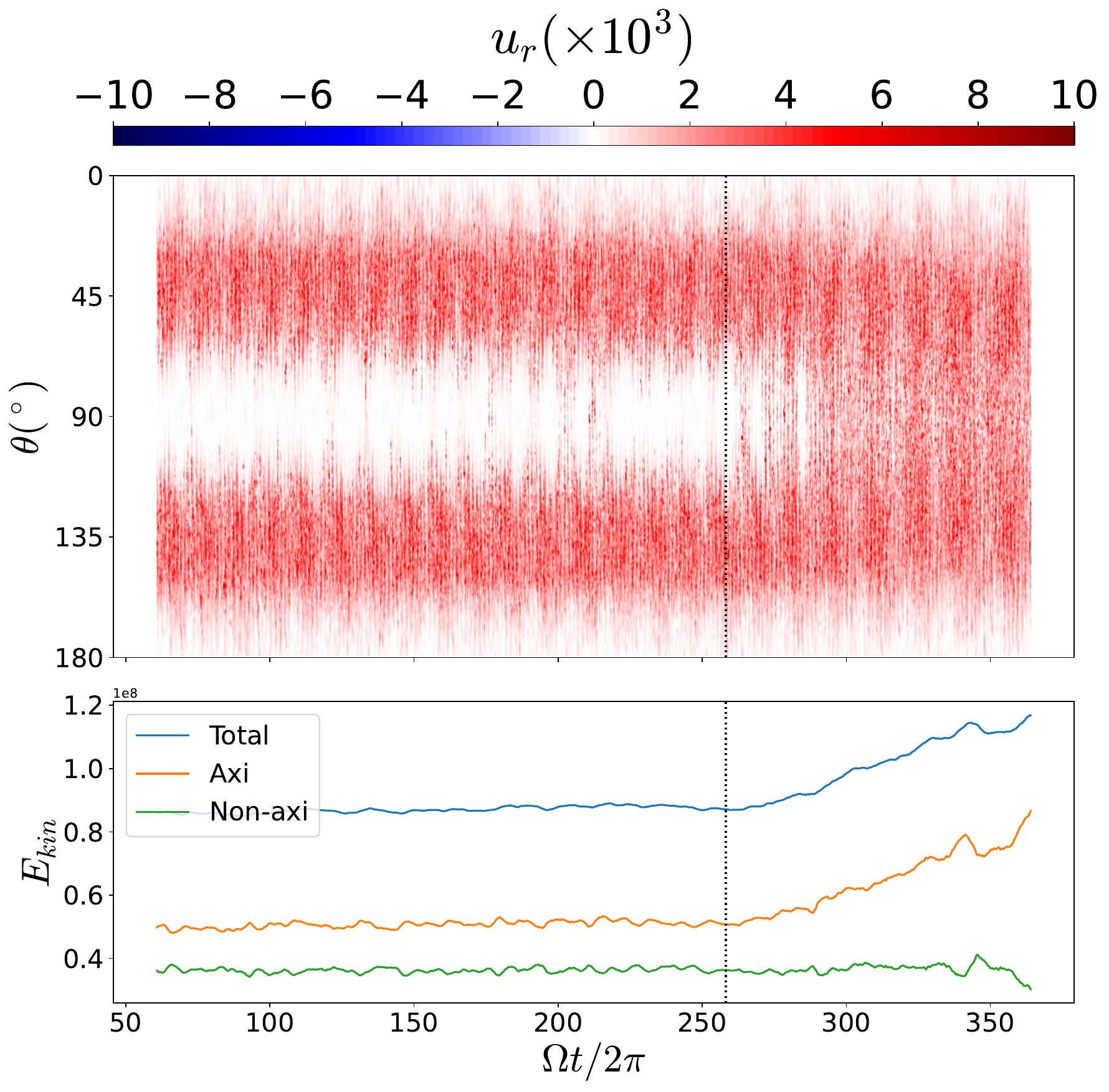}
    \caption{Transition to turbulence through instability at the equator at $E=3\times 10^{-5}, \Ro = -2$. Top panel shows radial velocity near the inner boundary $r/r_o = 0.36$ as a function of time (on the horizontal axis) and co-latitude $\theta$ on the vertical axis. Bottom panel shows the total, axisymmetric and non-axisymmetric kinetic energy as a function of time. The vertical dotted line marks the spin-up time based on $\Delta/\Omega$.}
    \label{fig:butterfly}
\end{figure}

As noted in section \ref{sec:spectra}, as $\Ro$ becomes increasingly negative, the flow near the inner boundary first becomes unstable at high latitudes and gives rise to small scale flows. At the transition to bulk turbulence, the flow near the inner boundary at and around the equator becomes unstable. This is illustrated in figure \ref{fig:vr3e-5}. Panel (a) shows radial velocity near the inner boundary before the transition to turbulence for suite S4 at $\Ro = -1.98$ and (b) shows the same after the transition to turbulence at $\Ro = -2$. The presence of small scales at all latitudes is markedly visible in panel (b). This is made even more clear if during this transition to turbulence, we track the radial velocity near the inner boundary at all latitudes and at a single longitude with respect to time. As illustrated in figure \ref{fig:butterfly}, when the fluid near the inner boundary spins up to $\Ro = -2$, small scale turbulent features start appearing near the equator. At the same time, the total and axisymmetric kinetic energies see a marked increase, as also explained in section \ref{subsec:MeanFlow}.

In order to visualise the dynamics of these small scales, we also produced a movie using snapshots of the simulation suite S3 at $E=10^{-4}, \Ro = -2.4$, in the turbulent regime. The initial condition was a solution at $\Ro = -2.25$, without any boundary layer instability at the equator. Several snapshots at regular intervals were used to produce the movie, which is available as a supplementary material (section \ref{SupMat}). The movie illustrates how small scale structures of high angular momentum fluid emanate from the equatorial boundary layer and give rise to a mean flow. It also illustrates in an equatorial section how the zonal flow is close to being axisymmetric and large scale initially, destabilising soon after as it transitions into the turbulent regime. As discussed in section \ref{subsec:MeanFlow}, a secondary pair of meridional circulation rolls are onset close to the inner boundary. However, the movie shows that their role is rather unimportant at this stage and the primary circulation is still responsible for most of the transport.

\begin{figure}
    \centering
    \includegraphics[width=\textwidth]{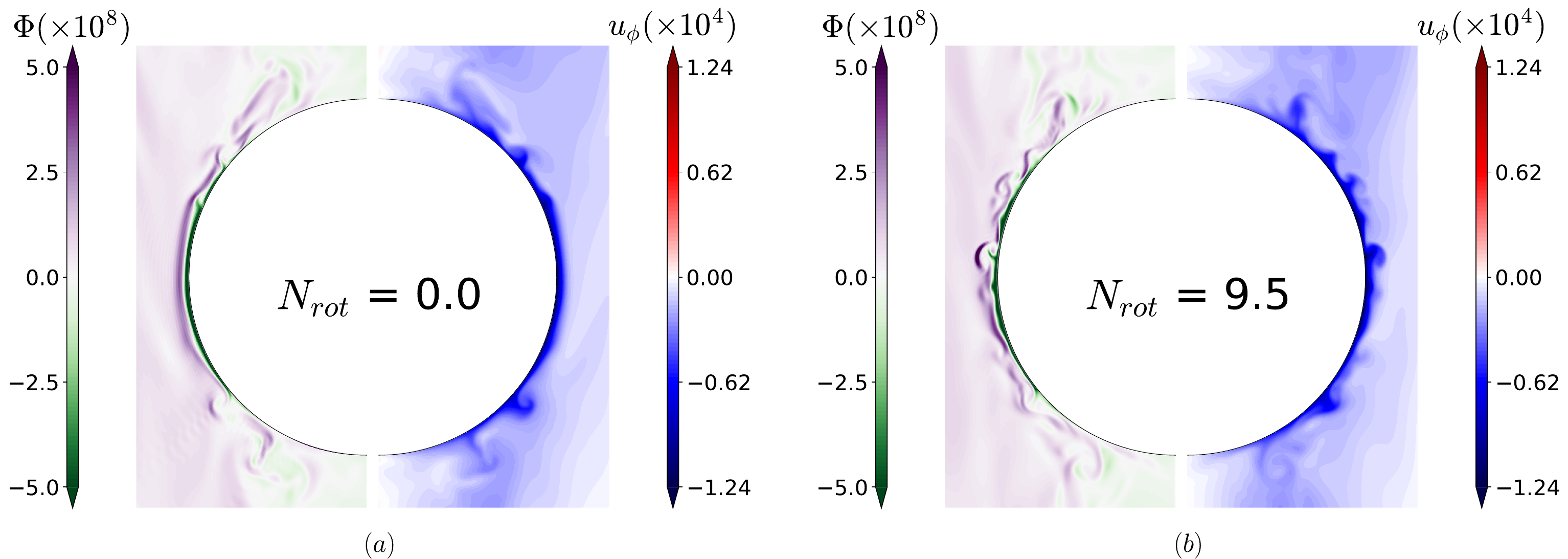}
    \caption{Visualising the Rayleigh stability criterion near the inner boundary for a couple of snapshots at $E=10^{-4}$ and $\Ro = -2.3$. The right panel in each case shows the zonal flow $u_\phi$ and the left panel shows the Rayleigh discriminant $\Phi$. $N_{rot}$ denotes the number of outer boundary rotations. (a) shows the state at the beginning of the simulation while (b) shows it after 9.5 rotations of the outer boundary (or 4 rotations of the inner boundary).}
    \label{fig:raycrit}
\end{figure}

The Rayleigh stability criterion in a rotating frame is given by \citep{Rayleigh1917, Kloosterziel91,Ghasemi2016},

\begin{equation}
    \Phi = \dfrac{\p}{\p s}(u_\phi s + \Omega s^2)^2 < 0\, ,
\end{equation}
where $\Omega = 1/E$ is the outer boundary rotation rate. We use a couple of snapshots in time of the zonal flow at zero longitude to visualise the Rayleigh discriminant $\Phi$ near the inner boundary. We use a simulation at $E=10^{-4}, \Ro=-2.3$ for this purpose. Figure \ref{fig:raycrit}(a) shows the at the beginning of the simulation when turbulence has not yet set in, while (b) shows the case after 9.5 outer boundary rotations when the boundary layer is fully unstable at all latitudes. We can see that $\Phi$ is strongly negative close to the inner boundary right at the start of the simulation. Though the colour-bars are the same for both plots, in terms of actual values of $\Phi$, at the beginning of the simulation (panel (a)), $\Phi_{min} = -2.02\times 10^{10}$ and $\Phi_{max} = 4.63\times 10^8$ while after 9.5 outer boundary rotations (panel (b)), $\Phi_{min} = -4.42\times 10^{9}$ and $\Phi_{max} = 7.28\times 10^8$. This implies that, in terms of extreme values of $\Phi$, the fluid near the inner boundary is 4.6 times more stable and only about half as unstable in panel (a) compared to panel (b). 
The boundary between strongly negative and positive parts correlate quite well with the unstable regions in panel (b).

\begin{figure}
    \centering
    \includegraphics[width=0.6\tw]{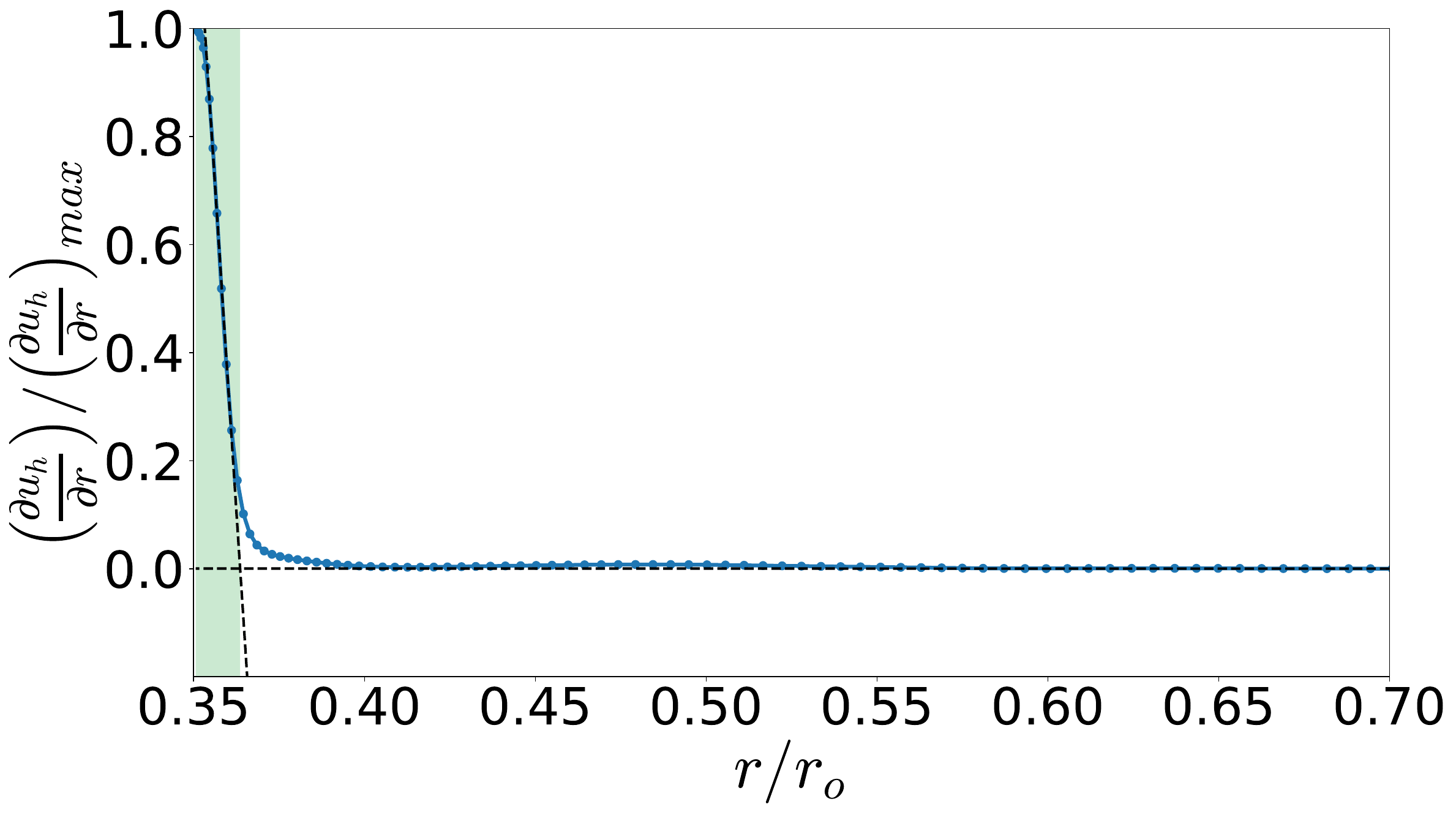}
    \caption{Illustration of how the thickness of the boundary layer at the inner boundary is determined by slope intersection method, at $E=3\times 10^{-5}, \Ro = -1.6$. The dashed lines show the slopes near the inner boundary and in the bulk, while the shaded region shows the boundary layer.}
    \label{fig:blThickIllus}
\end{figure}

\begin{figure}
    \centering
    \includegraphics[width=\textwidth]{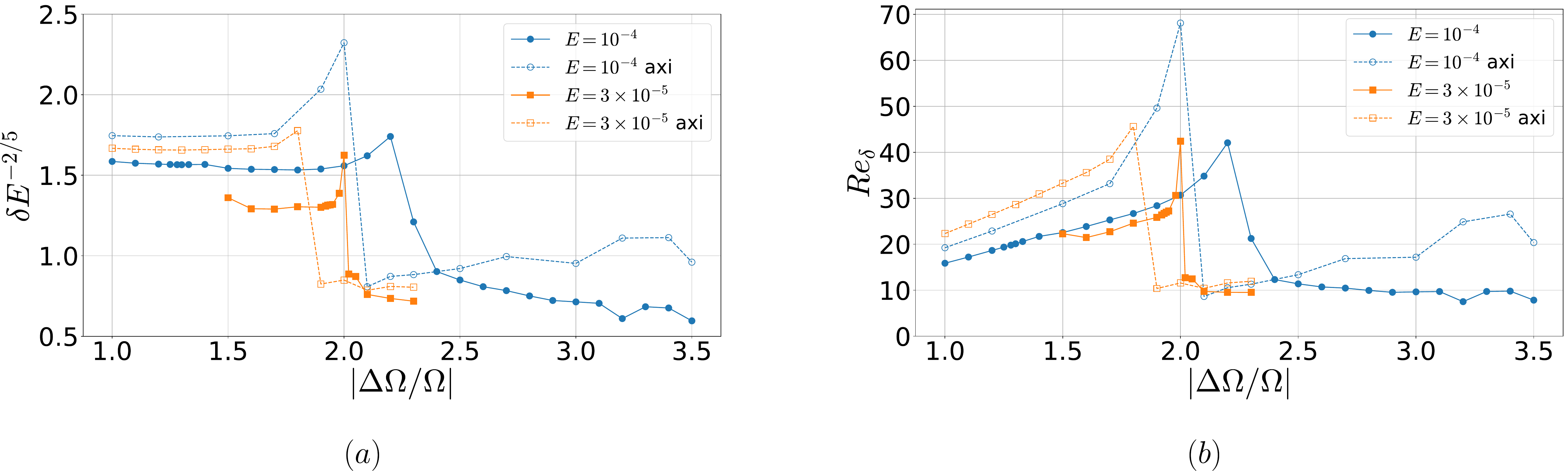}
    \caption{(a) Scaled thickness of the equatorial boundary layer $\delta$ as a function of $|\Ro|$ for 3D (solid lines, filled symbols) and axisymmetric simulations (dashed lines, open symbols) (b) $Re_\delta$ as a function of $\Ro$ for 3D simulation suites S3 and S4.}
    \label{fig:deltaBL}
\end{figure}

\begin{figure}
    \centering
    \includegraphics[width=0.6\tw]{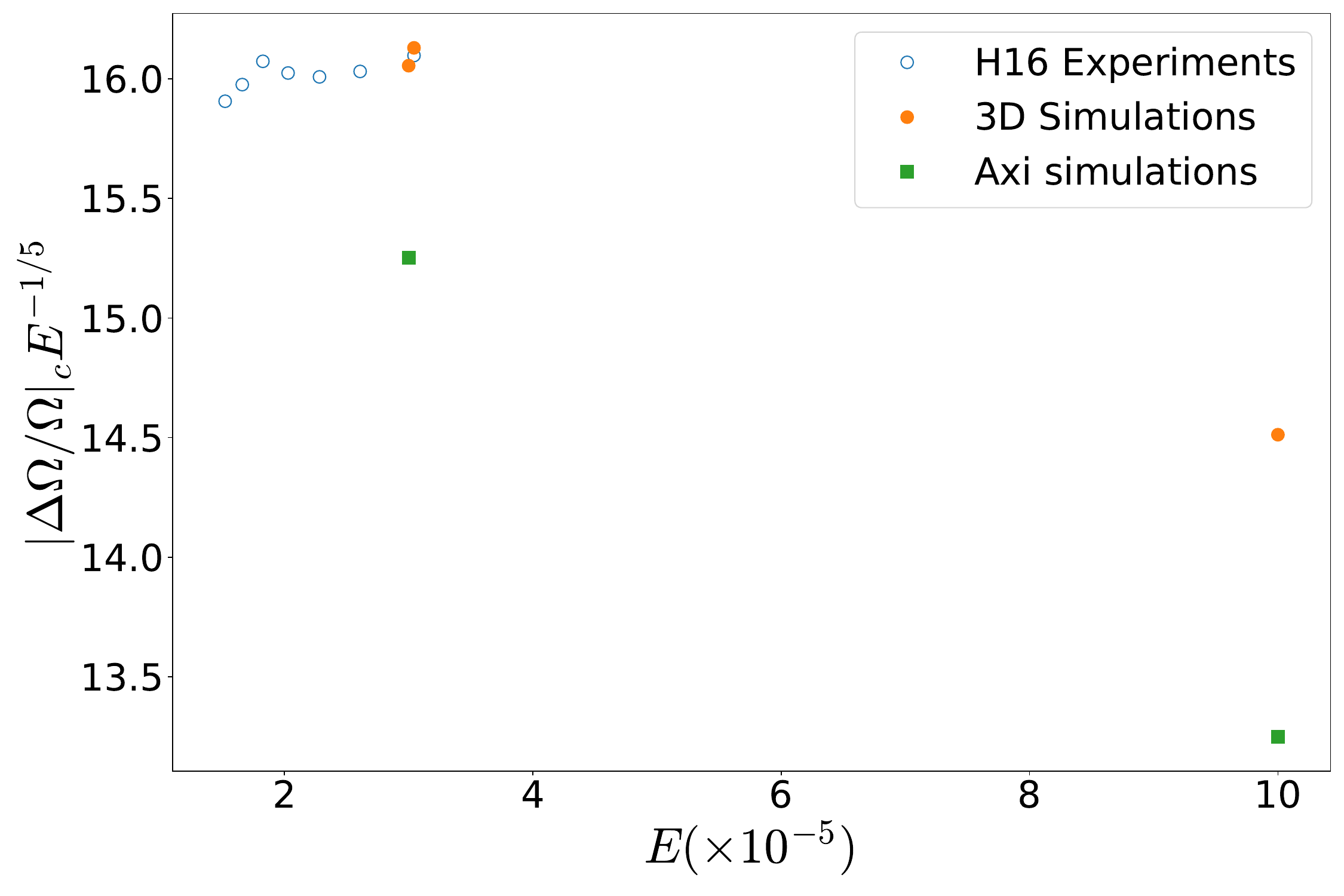}
    \caption{A compensated plot of $|\Ro|_c E^{-1/5}$ vs Ekman number for the experimental data of H16 (open circles), our 3D simulations (orange filled circles) and axisymmetric ones (green squares).}
    \label{fig:roScaling}
\end{figure}

We compute the thickness of the equatorial boundary layer at the inner boundary using a slope intersection method \citep[similar to e.g.][]{Verzicco99, Gastine2015,Barik2023}. We use time-averaged profiles of $\partial u_h/\partial r$, where $u_h = \sqrt{u_\theta^2 + u_\phi^2}$ is the magnitude of horizontal velocity. These profiles are obtained by averaging $u_h$ in azimuth and then in co-latitude with a window of $10^\circ$ centred at the equator. Fitting two lines to the respective profiles, one close to the inner boundary and a second one for the bulk, we assume that the boundary layer ends where both lines intersect. This is illustrated in figure \ref{fig:blThickIllus}. We then explore how the equatorial boundary layer thickness $\delta$ scales with differential rotation $|\Ro|$. The thickness is compensated by the theoretical $E^{2/5}$ scaling of the equatorial boundary layer thickness \citep{Stewartson66,Marcotte2016} in figure \ref{fig:deltaBL}(a). We can see that the scaling works rather well, except for the axisymmetric suite S3a near the transition to turbulence. The equatorial boundary layer thickness increases very slowly with $|\Ro|$ before the onset of turbulence. Close to the onset, there is an increase in the boundary layer thickness. After the transition to turbulence, the averaging in azimuth and time measures the thickness of the viscous sublayer, which is thinner than the laminar boundary layer and it decreases with Reynolds number \citep{LandauLif, Grossmann_Lohse_2000}. This is seen as a rapid decrease in $\delta$ beyond the transition and then a slow decrease with $|\Ro|$. We can define a Reynolds number based on the boundary layer thickness,

\begin{equation}
    Re_\delta = \dfrac{(\Omega_i - \Omega_o) \delta^2}{\nu} = \dfrac{(\Omega_i - \Omega_o)}{\Omega_o} \dfrac{\Omega_o L^2}{\nu} \left(\dfrac{\delta}{L}\right)^2 = \dfrac{\Delta\Omega}{\Omega} \dfrac{1}{E} \left(\dfrac{\delta}{L}\right)^2\, .
\end{equation}
If we assume that the boundary layer becomes turbulent once it exceeds a critical Reynolds number $Re_c$ and use the fact that $\delta/L = C E^{2/5}$, where $C$ is a constant, we find that at criticality,

\begin{equation}
\begin{split}
    C \left(\dfrac{\Delta\Omega}{\Omega}\right)_c \dfrac{1}{E} E^{4/5} &= Re_c~,\\
\Rightarrow \left(\dfrac{\Delta\Omega}{\Omega}\right)_c E^{-1/5} &= Re_c/C~.
\end{split}
\end{equation}
Figure \ref{fig:deltaBL}(b) shows the variation of $Re_\delta$ with $|\Ro|$ for all simulation suites. We find that, except for suite S3a, the rest of them peak at $Re_c = 42, 42, 45$ for suites S3, S4 and S4a, respectively. This implies that the assumption the existence of a critical Reynolds number works fairly well. Furthermore, figure \ref{fig:roScaling} shows the compensated plot of $|\Ro|_c E^{-1/5}$ with data from both experiments of H16 as well as our simulations. We find that the spread in the compensated plot is small, especially noting the variation along the vertical axis. The higher Ekman number simulations are slightly off a flat line, with the axisymmetric suite S3a being a complete outlier.

\section{Conclusion}\label{sec:conclusions}

The two sets of simulations at Ekman numbers of $10^{-4}$ and $3\times 10^{-5}$ presented here yield similar results and reproduce the experimental observations of \cite{Hoff2016} in the turbulent regime. These include the generation of zonal flow in the bulk, violating the classic solution for spherical Couette flow by \cite{Proudman56}, the loss of energy in inertial modes and inertial wave turbulence. Unfortunately, the experimental data is extremely limited in spatial extent, being limited to a single plane perpendicular to the rotation axis just above the inner sphere. This makes it difficult to make more quantitative comparisons with experiments beyond what we already made in \cite{Barik2018} and in section \ref{subsec:MeanFlow}. However, using simulations, we have been able to generate a more complete picture of the transition to turbulence. The cause of the onset of turbulence seems to be a centrifugal instability of the boundary layer at the equator of the inner boundary, giving rise to Taylor-G\"ortler vortices, similar to those observed by \cite{Noir2009} and \cite{Ghasemi2016, Ghasemi2018}. The hysteresis exhibited by the system \citep{EgbersRath95} implies a subcritical transition. Beyond the regimes of axisymmetric flow and the first linear instability, as the differential rotation rate is made increasingly negative, we find that the boundary layer at the inner boundary first becomes unstable at high latitudes. This is seen in both in spectral space (section \ref{sec:spectra}) as well as physical space (section \ref{sec:BL}). This instability gives place to spiral structures along the boundary layer, ejecting small scale plume-like structures. These small scale structures in a rotating environment are known to excite inertial waves \citep{Davidson2006, DavidsonTurbulence}, leading eventually to inertial wave turbulence as shown in section \ref{sec:spectra}.

At a critical negative differential rotation, the boundary layer at the inner boundary becomes unstable at the equator and thus, the resultant G\"ortler vortices can now propagate into the bulk, outside the tangent cylinder. They further contribute to an increase in the energy carried by inertial waves as well as to an increase in energy in the small scales away from the inner boundary as evidenced by the temporal and spatial spectra (section \ref{sec:spectra}). A significant increase in Reynolds stresses driving zonal flow ensues, which leads to the zonal flow spreading outside the tangent cylinder just as seen in experiments as well as simulations (section \ref{subsec:MeanFlow}). This also leads to a more efficient angular momentum transport, and thus to an increase in the scaling exponent of the torque at the inner boundary from linear to quadratic. A second set of axisymmetric simulations at $E=10^{-4}$ and $3\times 10^{-5}$ show a very similar behaviour in terms of creation of large scale zonal flow, torque scalings and destabilization of the inner boundary layer near the equator. However, in this case the instability of the equatorial boundary layer at the inner boundary gives rise to an equatorial jet, which makes the subsequent evolution of the centrifugal instability markedly different than the 3D simulations.
This equatorial jet also serves to transport angular momentum in the axisymmetric cases as opposed to Reynolds stresses for the full 3D simulations.

The Ekman layer near the inner boundary merges with the Stewartson layer into a layer that has an extent of $\delta_s \times \delta_z = E^{2/5}\times E^{1/5}$ \citep{Stewartson66,Marcotte2016} with $s$ and $z$ representing the cylindrical radius and axial direction, respectively. Using a heuristic critical Reynolds number argument for the destabilisation of the equatorial boundary layer at the inner boundary, we show that this scaling can help explain the experimental $E^{1/5}$ scaling for the critical differential rotation, especially at lower Ekman numbers (section \ref{sec:BL}). The finer details of this transition are something that can still be explored and investigated, but the centrifugal instability of the equatorial boundary layer is the clear precursor. It remains to be seen whether this scaling law extends to asymptotically low Ekman numbers. The narrowing gaps between the values of $(\Ro)_c$ for the full 3D and the axisymmetric simulations and their similar nature of instabilities is encouraging. This could enable us to obtain an estimate of $(\Ro)_c$ at lower Ekman numbers with cheaper axisymmetric computations.

Furthermore, our previous \citep{Barik2018} and current study has been limited to $\Ro < 0$. An in-depth study for $\Ro > 0$ is still lacking. In particular, it is not clear why one obtains high wavenumber spiral Stewartson layer instabilities for $\Ro > 0$, but low wavenumber instabilities trapped inside TC for $\Ro < 0$ \citep{Hollerbach2003, WichtJFM}. More simulations and experiments are needed to establish better scaling laws pertaining to the different hydrodynamic regimes at lower Ekman numbers. This will prove helpful not only in order to extrapolate to real objects, but also to understand the dichotomies between $\Ro < 0$ and $\Ro > 0$. The theoretical foundation for spherical Couette flow is still in its infancy as compared to the more traditional Taylor-Couette system \citep{GrossmannEtAl2016}. Our present study shows that there is a great scope for similar studies in spherical shells as well, where the presence of spherical curvature makes the problem less tractable.

\backsection[Supplementary data]{\label{SupMat}} The movie for transition to turbulence can be found at \url{doi.org/10.6084/m9.figshare.9108533}. The full 4k version can be viewed as an unlisted youtube video: \url{https://youtu.be/6vBWwYIapC8}.

\backsection[Acknowledgements]{A.B would like to thank Andreas Tilgner, Jonathan Aurnou, Nathana\"el Schaeffer and Paula Wulff  for insightful discussions and Sabine Stanley for feedback on the manuscript draft. We gratefully acknowledge Adrian Mazilu from the Transylvania University of Brasov (Romania) for performing most of the experiments in the frame of a Traineeship ERASMUS+ program.}

\backsection[Funding]{A.B would like to thank the IMPRS for Solar System Science for funding him during his time in Germany and Sabine Stanley for funding him subsequently. A.B would also like to thank the North-German Supercomputing Alliance (HLRN), Max Planck Computing and Data Facility (MPCDF) and the Gesellschaft f\"{u}r wissenschaftliche Datenverarbeitung mbH G\"{o}ttingen (GWDG) for letting him generously use their supercomputing facilities. S.A.T would like to acknowledge support from the European Research Council (ERC Advanced Grant 670874 ROTANUT) and from the infrastructure program EuHIT of the European Commission. M.H gratefully acknowledges support from the German Research Foundation (DFG Grant No. HA 2932/7-1).}

\backsection[Declaration of interests]{The authors report no conflict of interest.}

\backsection[Data availability statement]{Both codes used to run the simulations listed here are openly available. MagIC is available at \url{https://github.com/magic-sph/magic} and XSHELLS at \url{https://bitbucket.org/nschaeff/xshells}. The parameters used to run the codes are available in the paper and in \cite{Barik2018}.}

\backsection[Author ORCIDs]{A. Barik, https://orcid.org/0000-0001-5747-669X; S. A. Triana, https://orcid.org/0000-0002-7679-3962; J. Wicht, https://orcid.org/0000-0002-2440-5091 }

\backsection[Author contributions]{Ankit Barik ran the simulations and performed the subsequent data analysis and wrote the first draft of the manuscript. Santiago Triana ran the XSHELLS simulations and generated the resultant spectrogram. Both Santiago Triana and Michael Hoff participated in running the experiments at BTU-CS, Cottbus and the experimental data analysis. Michael Hoff performed a large part of the post processing of the experimental data. Johannes Wicht supervised the project and provided key insights. All authors contributed to providing feedback and refining the manuscript into its present form.}

\bibliographystyle{jfm}
\bibliography{Bibliography}

\end{document}